\documentclass[12pt,a4paper]{article}
\usepackage{amsmath,microtype}
\usepackage{graphicx,psfrag,epsf,orcidlink,setspace}
\usepackage{enumerate}
\usepackage[font=small]{caption}
\usepackage[round,authoryear]{natbib}
\usepackage{url, paralist} 
\usepackage[top=1.6in, bottom=1.6in, left=1.1in, right=1.1in]{geometry}
\usepackage{float,bm,booktabs,makecell,amsthm,amssymb,amsfonts,dsfont,mathtools,authblk,subcaption}
\newcommand{\commHS}[1]{{\leavevmode\color{purple}#1}}

\DeclareMathOperator*{\argmin}{arg\,min}

\usepackage{bbold}
\usepackage{algorithm}
\usepackage{algorithmic}

\usepackage[font=small]{caption}
\graphicspath{{plots/}}

\usepackage{amssymb}
\usepackage{hyperref}
\usepackage{multirow}
\usepackage{booktabs}
\usepackage{mathrsfs, xcolor} 
\definecolor{darkblue}{rgb}{0,0,.6}
\hypersetup{colorlinks = true, linkcolor=darkblue, urlcolor=darkblue, citecolor=darkblue}
\usepackage{xr}

\newcommand{\blind}{0}

\date{}

\begin{document}

\def\spacingset#1{\renewcommand{\baselinestretch}%
{#1}\small\normalsize} \spacingset{1}

\def\be{\begin{equation}}
\def\ee{\end{equation}} 
\def\ben{\begin{equation*}}
\def\een{\end{equation*}}
\def\bea{\begin{eqnarray}}
\def\eea{\end{eqnarray}}
\def\bda{\begin{eqnarray*}}
\def\eda{\end{eqnarray*}}
\numberwithin{equation}{section}

\newtheorem{definition}{Definition}
\newtheorem{theorem}{Theorem}
\newtheorem{proposition}{Proposition}
\newtheorem{corollary}{Corollary}
\newtheorem{assumption}{Assumption}
\renewcommand\theassumption{A\arabic{assumption}}
\newtheorem{lemma}{Lemma}
\newtheorem*{remark}{Remark}

\theoremstyle{definition}
\newtheorem{exmp}{Example}[section]
\AtEndDocument{\refstepcounter{theorem}\label{finalthm}}
\AtEndDocument{\refstepcounter{proposition}\label{finalprop}}
\newcommand{\pkg}[1]{{\normalfont\fontseries{b}\selectfont #1}} \let\proglang=\textsf \let\code=\texttt


\if0\blind
{
  \title{\bf Spherical Spatial Autoregressive Model for Spherically Embedded Spatial Data}
  \author{Jiazhen Xu \orcidlink{0009-0006-7870-0340}\footnote{Corresponding author: Department of Actuarial Studies and Business Analytics, Level 7, 4 Eastern Road, Macquarie University; Email: jiazhen.xu@mq.edu.au} \qquad Han Lin Shang \orcidlink{0000-0003-1769-6430} \\
  Department of Actuarial Studies and Business Analytics \\
  Macquarie University, 2113, NSW, Australia}
  \maketitle
} \fi

\if1\blind
{
  \bigskip
  \bigskip
  \bigskip
  \begin{center}
    {\LARGE\bf Spherical Spatial Autoregressive Model for Spherically Embedded Spatial Data}
\end{center}
  \medskip
} \fi

\bigskip

\begin{abstract}

Spherically embedded spatial data are spatially indexed observations whose values naturally reside on or can be equivalently mapped to the unit sphere. Such data are increasingly ubiquitous in fields ranging from geochemistry to demography. However, analysing such data presents unique difficulties due to the intrinsic non-Euclidean nature of the sphere, and rigorous methodologies for statistical modelling, inference, and uncertainty quantification remain limited. This paper introduces a unified framework to address these three limitations for spherically embedded spatial data. We first propose a novel spherical spatial autoregressive model that leverages optimal transport geometry and then extend it to accommodate exogenous covariates. Second, for either scenario with or without covariates, we establish the asymptotic properties of the estimators and derive a distribution-free Wald test for spatial dependence, complemented by a bootstrap procedure to enhance finite-sample performance. Third, we contribute a novel approach to uncertainty quantification by developing a conformal prediction procedure specifically tailored to spherically embedded spatial data. The practical utility of these methodological advances is illustrated through extensive simulations and applications to Spanish geochemical compositions and Japanese age-at-death mortality distributions.

\end{abstract}
\noindent%

\newpage
\spacingset{1.8}

\section{Introduction}

With advancements in data collection technologies, the acquisition of various types of object oriented data has become increasingly common and accessible. Beyond the standard case of points in Euclidean space, object oriented data encompass shapes, points on curved manifolds, phylogenetic trees, sound recordings, and images \citetext{see, e.g., \citealp{nye2011principal, srivastava2016functional, petersen2019frechet, scealy2023score}}. Departing from traditional statistical approaches that assume vector-valued data, the central principle of object oriented data analysis is to treat each data object as the fundamental unit of analysis \citep{marron2021object}. These objects often reside in general metric spaces, which can be non-Euclidean. Unlike Euclidean spaces, non-Euclidean metric spaces typically do not support operations like addition, scalar multiplication, or inner products, necessitating statistical methodologies that do not rely on Euclidean structure.

A particularly important class of non-Euclidean objects consists of those that are naturally, or can be equivalently, represented as points on a sphere whose dimension may be infinite, such as directions, compositional data, and distributional data. These data types, which we refer to as spherically embedded data, arise frequently across a wide range of scientific applications and are often collected over spatial locations. Important examples include geochemical compositions and life-table death counts \citetext{see, e.g., \citealp{aitchison1984statistical, rollinson2014using, shang2017grouped}}.

\begin{figure}[!htb]
\centering
\includegraphics[scale=0.22]{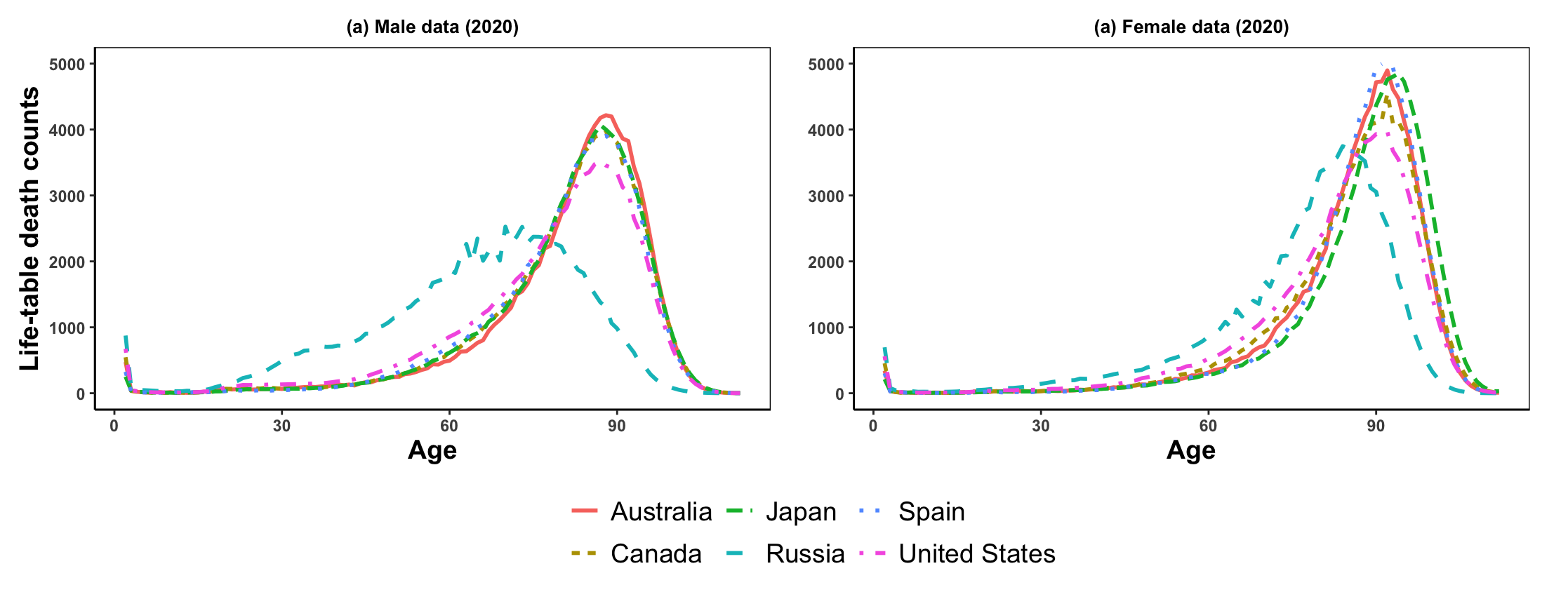}
\caption{Plots of the age distribution of deaths in 2020, by single-year age group, for (a) males and (b) females across six countries.}\label{fig::world LFDT}
\end{figure}

Despite the well-established body of work on Euclidean-valued spatial data \citep{cressie2015statistics,schabenberger2017statistical}, research on spherically embedded spatial data that does not rely on projecting the data into Euclidean space remains limited. \citet{xu2025generalized} propose a parametric model for finite-dimensional spherical spatial data under a specific distributional assumption, which cannot easily be extended to infinite-dimensional spherical spatial data. \citet{hoshino2024functional} considers distributional spatial data and first transforms the distributions into quantile functions, enabling subsequent methods to operate in a vector space. However, some distributional data do not have a one-to-one correspondence with quantile functions when bathtub-shaped hazards are present. For example, the age distribution of death often exhibits a pronounced concentration in the first year of life, as illustrated in Figure~\ref{fig::world LFDT} on age distributions across six countries.

Motivated by these challenges, this paper introduces a novel framework for spherically embedded spatial data that does not assume any parametric distribution. Our framework consists of three main components: 
\begin{inparaenum}
\item[(i)]  statistical modelling for spherically embedded spatial data, 
\item[(ii)] statistical inference for assessing spatial dependence, and \item[(iii)] a distribution-free prediction procedure that quantifies predictive uncertainty. 
\end{inparaenum}
Each component is of interest in its own right, is new to the literature, and is specifically designed to address the challenges arising from the non-Euclidean structure of spherically embedded spatial data.

For statistical modelling, we develop a novel spatial autoregressive model based on optimal transport for spherically embedded data, termed the spherical spatial autoregressive (SSAR) model. This approach extends the autoregressive (AR) model for spherical time series introduced in \cite{zhu2024spherical} to spatial contexts, allowing each observation to depend on neighbouring observations via a spatial weight matrix. In this sense, our method can be viewed as a generalisation of the spherical AR model in \cite{zhu2024spherical}. While several recent works have proposed autoregressive structures for distributional data using the optimal transport based on the Wasserstein metric \citep[e.g.,][]{jiang2022wasserstein, zhang2022wasserstein, chen2023wasserstein, zhu2023autoregressive, ghodrati2024distributional}, there are two key advantages of using optimal transport for spherical data. First, the spherical-based optimal transport works with a wider range of object types, such as spherical, compositional, and distributional data, whereas the Wasserstein-based optimal transport is primarily designed for distributional data. Second, it is easier to deal with multivariate distributional data using spherical-based optimal transport than using Wasserstein-based optimal transport. A more detailed comparison between the spherical-based optimal transport and the Wasserstein-based optimal transport can be found in \citet{zhu2024spherical}.

Notably, the AR modelling in \cite{zhu2024spherical} assumes that the Fr\'{e}chet mean of the spherically embedded observations is constant, and it is an assumption that becomes restrictive in spatial settings. Our SSAR model relaxes this constraint by allowing a varying Fr\'{e}chet mean through the incorporation of exogenous covariates. Methodologically, the SSAR model-fitting approach departs substantially from the existing spherical AR modelling method in \cite{zhu2024spherical}. In contrast to the Yule--Walker-type estimation of the AR coefficient used in \citet{zhu2024spherical}, the spatial autoregressive coefficient in the SSAR model requires a different estimation strategy due to the spatial dependence structure. Furthermore, the asymptotic theory in \citet{zhu2024spherical} assumes the constant population Fr\'{e}chet mean is known, whereas practical applications typically rely on the empirical Fr\'{e}chet mean. Our theoretical development explicitly accommodates this empirical mean, enabling a more realistic and broadly applicable asymptotic analysis.

Regarding statistical inference, we develop a Wald test for the spatial parameter introduced in our autoregressive model. To date, there are few methods for assessing spatial dependence in spherically embedded data without transformation. For example, \citet{xu2025generalized} develops a test for spherical spatial data based on the von Mises--Fisher distribution, which cannot be extended to the Hilbert sphere with infinite dimension. In contrast, our test does not require any distributional assumptions and can handle the Hilbert sphere case. Notably, the proposed Wald test statistic based on the asymptotic distribution of the spatial effect parameter estimators involves an estimator of the sample covariance (operator), and this sample estimator may suffer from the curse of dimensionality for small sample sizes when the dimension of the sphere is large or infinite. To tackle this issue, we further develop a bootstrap-based procedure for the Wald test to improve the finite-sample performance in practice.

Quantifying predictive uncertainty is also a critical task for spherically embedded spatial data. We develop a prediction set inspired by the conformal prediction framework (\citealt{SV08, FZV23}), ensuring validity without any distributional assumptions. While conformal prediction has been extended to non-Euclidean random objects in \cite{zhou2025conformal}, the existing method relies on the independent and identically distributed (IID) assumption, making it unsuitable for spherically embedded spatial data with spatial dependence. Our approach addresses this gap by providing a distribution-free prediction procedure that accommodates the spatial structure of spherically embedded data.

Our proposed approach represents a significant advancement in the analysis of spherically embedded spatial data. To the best of our knowledge, this is the first general framework for this problem that encompasses:
\begin{itemize}
\vspace{-0.8em}
\item  A spatial autoregressive model for spherically embedded spatial data, with an extension to include exogenous variables. We develop a parameter estimation approach based on the generalised method of moments (GMM), along with theoretical results on the asymptotic behaviour of the estimators.
\vspace{-0.8em}
\item A Wald test procedure derived from the central limit theorem for the parameter estimates and a bootstrap-based test to improve finite-sample performance. Both procedures are supported by theoretical guarantees.
\vspace{-0.8em}
\item A conformal inference methodology to quantify the uncertainty of predictions produced by the fitted spherical spatial autoregressive model, together with theoretical supports.
\vspace{-0.8em}
\item A demonstration of the practical merits of the proposed framework through comprehensive simulation studies and real data analyses on geochemical data and age distributions of death.
\end{itemize}

The remainder of the paper is organised as follows. Section~\ref{sect::spatial spherical data} introduces the formal definition of spherically embedded spatial data. Section~\ref{sect::PSSAR} presents the pure SSAR model without exogenous covariates, along with the associated parameter estimation procedure, hypothesis-testing framework, and conformal inference methodology, each with their theoretical guarantees. Section~\ref{sect::SRMSAR} extends the SSAR model to incorporate exogenous regressors and develops the corresponding estimation, testing, and conformal inference procedures. Section~\ref{sect::simulation} evaluates the finite-sample performance of the proposed methods through simulations under a variety of spatial dependence structures. Section~\ref{sect::real data} demonstrates the practical utility of our approach through applications to three real-world datasets. A conclusion and a discussion are given in Section \ref{sec:conclu}. Assumptions are provided in the Appendix, and all proofs are deferred to the Supplementary Material.

\section{Spherically Embedded Spatial Data}\label{sect::spatial spherical data}

Observations of spherically embedded spatial data are indexed by spatial locations and either arise naturally on a sphere or can be equivalently represented as elements of the sphere $\mathcal{S}=\{\nu\in\mathcal{H}:\|\nu\|_{\mathcal{H}}=1\}$. Here $\mathcal{H}$ is a separable Hilbert space with inner product $\langle\cdot,\cdot\rangle_{\mathcal{H}}$ and the Hilbert norm $\|\nu\|_{\mathcal{H}}=\sqrt{\langle\nu,\nu\rangle_{\mathcal{H}}}$. The metric space $\mathcal{S}$ is equipped with the intrinsic metric $d_\mathcal{S}(\nu_1,\nu_2)=\arccos(\langle\nu_1,\nu_2\rangle_\mathcal{H})$ for $\nu_1,\nu_2\in \mathcal{S}$. Notably, the sphere $\mathcal{S}$ covers both the finite-dimensional sphere and the infinite-dimensional sphere. If the sphere is finite-dimensional and is surrounded by a Hilbert space $\mathcal{H}=\mathbb{R}^m$ for an integer $m$, we further denote this finite-dimensional sphere as $\mathcal{S}^{m-1}$. If the sphere is infinite-dimensional, we further denote it by $\mathcal{S}^{\infty}$. 

Examples of random objects that can be equivalently represented as objects on a sphere include compositional data and distributional data. The space of compositional data is a simplex $\Delta^{m-1} = \Big\{ \bm{\delta}=(\delta_1,\delta_2,\ldots,\delta_m)^\top \in \mathbb{R}^m, \quad \sum_{l=1}^m \delta_l=1 ~{\rm and}~\delta_l\geq 0,\quad l=1,2,\ldots,m. \Big\}$. As studied in \cite{SW11}, a pointwise square root transformation $t_p:\Delta^{m-1}\mapsto \mathcal{S}^{m-1}$ given by $t_p(\bm{\delta})=(\sqrt{\delta_1},\sqrt{\delta_2},\ldots,\sqrt{\delta_m})^{\top}$ maps the objects in $\Delta^{m-1}$ to the finite sphere $\mathcal{S}^{m-1}$, where $^{\top}$ denotes the matrix transpose. For density functions $g:\mathbb{R}^l\mapsto \mathbb{R}$, one can use a functional pointwise square root transformation $t_f(g) = z,~{\rm where}~z(\omega)=\sqrt{g(\omega)}~{\rm for~all}~\omega\in\mathbb{R}^l$ to map the density function to the Hilbert sphere $\mathcal{S}^\infty$ equipped with the Fisher-Rao metric (\citealt{dai2022statistical,zhu2024spherical}).

\section{Pure Spherical Spatial Autoregressive Model}\label{sect::PSSAR}

To build the pure SSAR model without exogenous covariates, we first revisit the classical pure spatial autoregressive (SAR) model \citetext{see, e.g., \citealp{whittle1954stationary, mead1967mathematical}} in the Euclidean setting for real-valued observations $z_1,z_2,\ldots,z_n$ which takes the form of
\begin{align}\label{model::pure SAR process}
z_i=\rho_0\sum_{j=1}^n w_{i,j}z_j +\epsilon_i,~i=1,\ldots,n.
\end{align}
Here, $\rho_0$ represents the spatial effect due to the influence of neighbouring units on a single spatial unit. Under~\eqref{model::pure SAR process}, $z_i$ is assumed to have a constant zero mean across $i\in\{1,\ldots,n\}$. We can generalise~\eqref{model::pure SAR process} as
\begin{align}\label{model::pure SAR process ver2}
z_{i}-m_{c}=\rho_0\sum_{j=1}^n w_{i,j}(z_j-m_c) +\epsilon_i,\quad i=1,\ldots,n,
\end{align}
where $m_c=\mathbb{E}(z_1)$. Inspired by the idea from \cite{zhu2023autoregressive} and \cite{zhu2024spherical}, $z_i-m_c$ in~\eqref{model::pure SAR process ver2} can be interpreted as an optimal transport map that moves $m_c$ to $z_i$ in the Euclidean space. For spherical data $\nu_1,\nu_2\in \mathcal{S}$, \cite{zhu2024spherical} construct an optimal transport map that moves $\nu_2$ to $\nu_1$ as $\nu_2 \ominus \nu_1 =  T_{\nu_1,\nu_2}=\theta (\zeta_2\circ \zeta_1 - \zeta_1\circ\zeta_2)$, where $\theta=d_\mathcal{S}(\nu_1,\nu_2)$, $\zeta_1=\nu_1$, $\zeta_2=(\nu_2-\langle\nu_2,\nu_1\rangle_{\mathcal{H}}\nu_1)/\|\nu_2-\langle\nu_2,\nu_1\rangle_{\mathcal{H}}\nu_1\|_{\mathcal{H}}$, and $\circ$ denotes the tensor product. This construction ensures that the optimal transport map $T_{\nu_1,\nu_2}$ belongs to a set of skew-symmetric operators, denoted as $\mathcal{C}(\mathcal{H})$, which is Hilbertian with the Hilbert-Schmidt inner product as $\langle h_1\circ \widetilde{h}_1, h_2\circ \widetilde{h}_2 \rangle_{\mathcal{C}(\mathcal{H})} = \langle h_1 , h_2 \rangle_{\mathcal{H}} \langle \widetilde{h}_1 , \widetilde{h}_2 \rangle_{\mathcal{H}}$. Also, by the Rodrigues rotation formula in \citet[][Theorem~1]{zhu2024spherical}, one has the exponential of $T_{\nu_1,\nu_2}$ as
\begin{equation*}
\exp(T_{\nu_1,\nu_2}) = {\rm id}  + \frac{\sin(\theta)}{\theta}T_{\nu_1,\nu_2} + \frac{1-\cos(\theta)}{\theta^2}T_{\nu_1,\nu_2}^2,
\end{equation*}
for an angle $\theta\in[0,2\pi]$ and ${\rm id}$ denotes the identity operator. Based on the optimal transport map $T_{\nu_1,\nu_2}$, let the geodesic $\gamma:[0,1]\mapsto\mathcal{S}$ as $\gamma(\delta)=\exp (\delta T_{\nu_1,\nu_2})\nu_1$, we have $\gamma(0)=\nu_1$ and $\gamma(1)=\nu_2$, see the detailed deviation in Section~S1 of the Supplementary Material.

Following the Hilbertian property of the set of $T_{\nu_1,\nu_2}$ and recalling that the optimal transport $\nu_2 \ominus \nu_1$ is interpreted as the map that moves $\nu_1$ to $\nu_2$, given a sequence of spherical observations $y_1,y_2,\ldots,y_n\in\mathcal{S}$ with a constant Fr\'{e}chet mean $\mu_c=\argmin_{\nu\in\mathcal{S}} \mathbb{E}\{ d_\mathcal{S}^2 (\nu,y_1) \}$, the proposed pure spherical spatial autoregressive (PSSAR) model is 
\begin{align}\label{model::PSSAR}
q_i-\overline{q}=\rho_0\sum_{j=1}^n w_{i,j}(q_j-\overline{q})+\epsilon_i,~i=1,\ldots,n,
\end{align}
where $q_i=y_i \ominus \mu_c$, $\overline{q}=\mathbb{E}[q_1]$, $\rho_0$ is the true but unknown spatial autoregressive parameter, and $\{\epsilon_i\}_{i=1}^n\subset \mathcal{C}(\mathcal{H})$ are IID random noise with mean 0 and finite fourth moments, such as $\varpi_2=\mathbb{E}\left( \langle \epsilon_1,\epsilon_1\rangle_{\mathcal{C}(\mathcal{H})}^2 \right)<\infty$. The reason why $q_1,\ldots,q_n$ have the same mean lies in the construction of the optimal transport map $T_{\nu_1,\nu_2}$.

Note that, using the stacked form, the PSSAR model~\eqref{model::PSSAR} can be represented as
\begin{align}\label{model::PSSAR stack form}
Q_{n}=\rho_{0} \mathcal{W}_nQ_n +\Gamma_{n},
\end{align}
where $Q_n=(q_1-\overline{q},\ldots,q_n-\overline{q})\in \mathcal{C}^n(\mathcal{H})$, $\mathcal{C}^n(\mathcal{H})$ is the Cartesian product space of $n$ copies of $\mathcal{C}(\mathcal{H})$, $ \mathcal{W}_n=W_n \otimes {\rm id}$, $W_n=(w_{i,j})_{n\times n}$, $\otimes$ is the Kronecker product, and we have $(\mathcal{W}_nQ_n)_i=\sum_{j=1}^n w_{i,j}(q_j-\overline{q})$, $\Gamma_n=(\epsilon_1,\ldots,\epsilon_n)\in \mathcal{C}^n(\mathcal{H})$. 

\subsection{Estimation and Hypothesis Testing}\label{subsect::PSSAR est and test}

The condition that the PSSAR model~\eqref{model::PSSAR stack form} has an equilibrium solution is that $S_n(\rho_0)=I_n-\rho_0 W_n$ is invertible, where $I_n\in\mathbb{R}^{n\times n}$ is the identity matrix. The equilibrium solution of~\eqref{model::PSSAR stack form} is then in the form of $Q_n=(S_n^{-1}(\rho_0)\otimes {\rm id})\Gamma_n$. To estimate the parameter $\rho$, we consider a GMM borrowing ideas from \cite{lee2001generalized}. The proposition~\ref{prop::MoM construct} given below provides the foundation of the GMM estimation.

\begin{proposition}\label{prop::MoM construct}
When $S_n(\rho_0)=I_n-\rho_0 W_n$ is invertible, for any constant $P_n=(P_{ij})_{n\times n}\in\mathbb{R}^{n\times n}$ with ${\rm tr}(P_n)=0$, $\mathbb{E} \left[ \langle\left\{(P_nS_n(\rho_0))\otimes {\rm id} \right\} Q_n , \Gamma_n \rangle_{\mathcal{C}^n(\mathcal{H})} \right] = 0$, where $\langle(h_1,\ldots,h_n),(\widetilde{h}_1,\ldots,\widetilde{h}_n)\rangle_{\mathcal{C}^n(\mathcal{H})}=\sum_{i,j=1}^n \langle h_i,\widetilde{h}_j\rangle_{\mathcal{C}(\mathcal{H})}$
\end{proposition}

Proposition~\ref{prop::MoM construct} indicates that one can construct valid instrumental variables based on some specific matrix $P_n$ satisfying ${\rm tr}(P_n)=0$. Consider $\widehat{Q}_n=(\widehat{q}_1-\widehat{q},\ldots,\widehat{q}_n-\widehat{q})$ where $\widehat{q}_i=y_i\ominus \widehat{\mu}_c$, $\widehat{\mu}_c=\argmin_{\nu\in\mathcal{S}}n^{-1}\sum_{i=1}^n d_\mathcal{S}^2(\nu,y_i)$ for $i=1,\ldots, n$, and $\widehat{q}=n^{-1}\sum_{i=1}^{n} \widehat{q}_i$. The proposition~\ref{prop::MoM construct} shows that the GMM estimator $\widehat{\rho}$ can be solved from the equation $\langle\left\{(P_nS_n(\rho))\otimes {\rm id} \right\} \widehat{Q}_n , \left\{S_n(\rho)\otimes {\rm id} \right\}\widehat{Q}_n \rangle_{\mathcal{C}^n(\mathcal{H})}=0$, which is equivalent to ${\rm tr}\left\{ S_n^\top(\rho) P_n S_n(\rho) \widehat{G}_n \right\} = 0$, where the $(j,k)$\textsuperscript{th} element of $\widehat{G}_n$ is $\langle \widehat{q}_j-\widehat{q} , \widehat{q}_k-\widehat{q} \rangle_{\mathcal{C}(\mathcal{H})}$. Therefore, the GMM estimator $\widehat{\rho}$ is then given by
\begin{align}\label{formula::PSSAR rho est}
\widehat{\rho} = \argmin_{\rho} \left[ {\rm tr}\left\{ S_n^\top(\rho) P_n S_n(\rho) \widehat{G}_n \right\} \right]^2.
\end{align}
Consider the $(k,l)$\textsuperscript{th} elements of $W_n$ and $P_n$ are of order $O(\eta_n^{-1})$ uniformly in $k$ and $l$, as describe in Assumption~\ref{assump::1} in the Appendix, the asymptotic distribution of $\widehat{\rho}$ is given in the following theorem.

\begin{theorem}\label{thm::PSSAR rho est}
Under Assumptions~\ref{assump::1} and~\ref{assump::2} in the Appendix, assume that $\lim_{n\to\infty}n^{-1}\eta_n {\rm tr} \{(P_n+P_n^\top)W_nS_n^{-1}(\rho_0) \}\neq 0$. The estimator $\widehat{\rho}$ from (\ref{formula::PSSAR rho est}) has the asymptotic behaviour as $\sqrt{n\eta_n^{-1}} (\widehat{\rho}-\rho_0)\xrightarrow{d} N(0,\sigma_\rho^2)$, where 
\begin{equation*}
\sigma_{\rho}^2 = \lim_{n\to\infty}\left[ \frac{\varpi_4{\rm tr}\left\{ P_n(P_n+P_n^\top)\right\}+(\varpi_5-2\varpi_4-1)\sum_{i=1}^n P_{n,ii}^2}{\eta_nn^{-1}{\rm tr}^2\left\{ (P_n+P_n^\top)W_nS_n^{-1}(\rho_0) \right\}} \right],
\end{equation*}
$\varpi_1={\rm tr}(\Sigma_\epsilon)$, $\Sigma_\epsilon=\mathbb{E}( \epsilon_1\circ\epsilon_1 )$, $\varpi_2=\mathbb{E}\left( \langle \epsilon_1,\epsilon_1\rangle_{\mathcal{C}(\mathcal{H})}^2 \right)$, $\varpi_3={\rm tr}(\Sigma_\epsilon^2)$, $\varpi_4=\varpi_3/\varpi_1^2$, and $\varpi_5=\varpi_2/\varpi_1^2$. If we further have 
\begin{inparaenum}
\item[(i)] $P_{n,ii}=0$ for all $i=1,\ldots,n$, or 
\item[(ii)] $\lim_{n\to\infty}\eta_n\to\infty$,
\end{inparaenum}
\begin{equation*}
\sigma_{\rho}^2 = \lim_{n\to\infty}\left[ \frac{\varpi_4{\rm tr}\left\{ P_n(P_n+P_n^\top)\right\}}{\eta_nn^{-1}{\rm tr}^2\left\{ (P_n+P_n^\top)W_nS_n^{-1}(\rho_0) \right\}} \right].
\end{equation*}
\end{theorem}

After obtaining the parameter estimator $\widehat{\rho}$ and its asymptotic property in Theorem~\ref{thm::PSSAR rho est}, we next assess the presence of the spatial effect with the following null and alternative hypotheses
\begin{equation}\label{two sample hypothesis}
    {\rm H}_0: \rho=0 ~~{\rm vs.}~~{\rm H}_a:\rho\neq0.
\end{equation}
A Wald-type test statistic is proposed to test ${\rm H}_0$ as
\begin{equation}\label{formula::PSSAR wald test}
\widehat{T}_{w,\rho} = \frac{n\eta_n^{-1}\widehat{\rho}^2}{\widehat{\sigma}_\rho^2},
\end{equation}
where $\widehat{\sigma}_\rho^2$ is a consistent estimator of $\sigma_\rho^2$ in Theorem~\ref{thm::PSSAR rho est}, and takes the form of
\begin{equation*}
\widehat{\sigma}_\rho^2 = \frac{\widehat{\varpi}_4{\rm tr}\left\{ P_n(P_n+P_n^\top)\right\}+(\widehat{\varpi}_5-2\widehat{\varpi}_4-1)\sum_{i=1}^n P_{n,ii}^2}{\eta_nn^{-1}{\rm tr}^2\left\{ (P_n+P_n^\top)W_nS_n^{-1}(\widehat{\rho}) \right\}},
\end{equation*}
with $\widehat{\varpi}_1={\rm tr}(\widehat{\Sigma}_\epsilon)$, $\widehat{\Sigma}_\epsilon=n^{-1}\sum_{i=1}^n \widehat{\epsilon}_i\circ\widehat{\epsilon}_i$, $\widehat{\epsilon}_i=\widehat{q}_i-\widehat{q}-\widehat{\rho}\sum_{j=1}^n w_{i,j}(\widehat{q}_j-\widehat{q})$ $i=1,\ldots,n$, $\widehat{\varpi}_2=n^{-1}\sum_{i=1}^n \langle \widehat{\epsilon}_i,\widehat{\epsilon}_i\rangle_{\mathcal{C}(\mathcal{H})}^2$, $\widehat{\varpi}_3=\{n(n-1)\}^{-1}\sum_{i\neq j} \langle \widehat{\epsilon}_i ,\widehat{\epsilon}_j \rangle_{\mathcal{C}(\mathcal{H})}$, $\widehat{\varpi}_4=\widehat{\varpi}_3/\widehat{\varpi}_1^2$, and $\widehat{\varpi}_5=\widehat{\varpi}_2/\widehat{\varpi}_1^2$. If we further have (i) $P_{n,ii}=0$ for all $i=1,\ldots,n$, or (ii) $\lim_{n\to\infty}\eta_n\to\infty$,
\begin{equation*}
\widehat{\sigma}_\rho^2 = \left[ \frac{\widehat{\varpi}_4{\rm tr}\left\{ P_n(P_n+P_n^\top)\right\}}{\eta_nn^{-1}{\rm tr}^2\left\{ (P_n+P_n^\top)W_nS_n^{-1}(\widehat{\rho}) \right\}} \right].
\end{equation*}
The asymptotic distribution of the proposed Wald test statistic $\widehat{T}_{w,\rho}$ in (\ref{formula::PSSAR wald test}) is given below.
\begin{theorem}\label{thm::wald test}
Under Assumptions~\ref{assump::1} and~\ref{assump::2} in the Appendix, assume that $\lim_{n\to\infty}n^{-1}\eta_n {\rm tr} \{(P_n+P_n^\top)W_nS_n^{-1}(\rho_0)\}\neq ~0$. Then, under the null hypothesis ${\rm H}_0$ in (\ref{two sample hypothesis}), $\widehat{T}_{w,\rho}\xrightarrow{d}\chi^2(1)$ as $n\to\infty$.
\end{theorem}

Notably, the estimate $\widehat{\sigma}_\rho^2$ in $\widehat{T}_{w,\rho}$ may suffer the curse of dimensionality when the sample size $n$ is small. Observe that $\widehat{\sigma}_\rho^2$ involves ${\rm tr}(\widehat{\Sigma}_\epsilon)$ which is a sum of the eigenvalues of $\widehat{\Sigma}_\epsilon$. Thus, a method based on principal component analysis (PCA) can be used to handle the higher dimensionality of the sphere $\mathcal{S}$. In practice, one can select eigenvalues above a threshold, chosen by the fraction of variance explained.

However, the PCA-based method still may fail for a small sample size with the dimension of the sphere is large or infinite, as the eigenvalues of $\widehat{\Sigma}_\epsilon$ become much noisier, which has been observed in our numerical studies in Section~\ref{subsect::PSSAR est and test}, together with Section~S4 in the Supplementary Material. To improve the finite sample performance of the hypothesis testing, we consider a bootstrap-based procedure, which is summarised as follows:
\begin{itemize}
\vspace{-0.8em}
\item[(a)] For $i=1,2,\ldots,n$, calculate $\widehat{\epsilon}_i=\widehat{q}_i-\widehat{q}$;
\vspace{-0.8em}
\item[(b)] Draw a bootstrap sample $\{\widehat{\epsilon}_i^*\}_{i=1}^n$ from $\{\widehat{\epsilon}_i\}_{i=1}^n$ and get $\widehat{q}_i^*=\widehat{\epsilon}_i^*+\widehat{q}$;
\vspace{-0.8em}
\item[(c)] Calculate the bootstrap estimate $\widehat{\rho}^{(b)}$ based on $\{\widehat{q}_i^*\}_{i=1}^n$;
\vspace{-0.8em}
\item[(d)] Repeat (a)-(c) $B$ times to obtain the empirical distribution function of $\{\widehat{\rho}^{(b)}\}_{b=1}^B$. For a nominal level $\alpha$, if $\widehat{\rho}$ is located outside the $(1-\alpha)$-level confidence interval based on the empirical distribution function of $\{\widehat{\rho}^{(b)}\}_{b=1}^B$, denoted as $[l_{\alpha/2}^{\rm PSSAR},u_{1-\alpha/2}^{\rm PSSAR}]$, reject the null hypothesis ${\rm H}_0$.
\end{itemize}

The following theorem ensures the validity of the proposed bootstrap procedure.

\begin{theorem}\label{thm::PSSAR bootstrap}
Under Assumptions~\ref{assump::1} and~\ref{assump::2} in the Appendix, assume that $\lim_{n\to\infty}n^{-1}\eta_n {\rm tr} \{(P_n+P_n^\top)W_nS_n^{-1}(\rho_0)\}\neq 0$. Then under the null hypothesis ${\rm H}_0$ in (\ref{two sample hypothesis}), $\lim_{n \to \infty} \mathbb{P}\left( l_{\alpha/2}^{\rm PSSAR} \leq \widehat{\rho} \leq u_{1-\alpha/2}^{\rm PSSAR} \right) = 1-\alpha$.
\end{theorem}

\subsection{Prediction and Conformal Prediction Set}\label{subsect::PSSAR prediction}

For the PSSAR model in~\eqref{model::PSSAR} with observations $y_1,y_2,\ldots,y_n$ and the weight matrix $W_n$, given the estimated parameters $\widehat{\mu}_c$ $\widehat{q}$, and $\widehat{\rho}$, the prediction for a new skew-symmetric operator at site $n+1$ and the weight matrix $W_{n+1}^*$ containing the spatial information of the sites $\{1,2,\ldots,n+1\}$ is $\widehat{q}_{n+1}=\widehat{q} + \widehat{\rho} \sum_{j=1}^{n} w_{n+1,j}^* (\widehat{q}_j-\widehat{q})$, where $w_{i,j}^*$ is the $(i,j)$\textsuperscript{th} element of $W_{n+1}^*$ for $i,j\in\{1,2,\ldots,n+1\}$. The prediction for the spherical data $y_{n+1}$ is then 
\begin{align}\label{formula::PSSAR prediction}
\widehat{y}_{n+1} = \exp(\widehat{q}_{n+1})\widehat{\mu}_c
=\left[{\rm id}+\sin(\widehat{\theta}_{n+1})\widehat{q}_{n+1}+\{1-\cos(\widehat{\theta}_{n+1})\}\widehat{q}_{n+1}^2\right]\widehat{\mu}_c,
\end{align}
where $\widehat{\theta}_{n+1}=\|\widehat{q}_{n+1}\widehat{\mu}_c\|_\mathcal{H}$ and $\widehat{\mu}_c=\argmin_{\nu\in\mathcal{S}}n^{-1}\sum_{i=1}^nd_\mathcal{S}^2(\nu,y_i)$.

To quantify the prediction uncertainty, we use a split-conformal method to construct the prediction set. We first randomly split $\{y_i\}_{i=1}^n$ into a training set $\mathcal{Y}_{\rm train}$ and a calibration set $\mathcal{Y}_{\rm cal}$ with the corresponding weight matrices. Denote the index sets for the training and calibration as $\mathcal{I}_{\rm train}$ and $\mathcal{I}_{\rm cal}$, respectively. For the training set, we train the PSSAR model in (\ref{model::PSSAR}). That is, the PSSAR model in (\ref{model::PSSAR}) is fitted using $\{\widehat{q}_k^{\rm train}=y_k\ominus\widehat{\mu}_c^{\rm train}:k\in \mathcal{I}_{\rm train}\}$ where $\widehat{\mu}_c^{\rm train}=\argmin_{\nu\in\mathcal{S}}\{{\rm card}(\mathcal{I}_{\rm train})\}^{-1}\sum_{k\in \mathcal{I}_{\rm train}}d_\mathcal{S}^2(\nu,y_k)$ with ${\rm card}(\mathcal{A})$ being the cardinality of a set $\mathcal{A}$, and the estimated spatial parameter is denoted as $\widehat{\rho}_{\rm train}$. The non-conformity score is calculated as $\widehat{r}_i=\|\widehat{\epsilon}_i^{\rm cal}\|_{\mathcal{C} (\mathcal{H})}$ where for $i\in\mathcal{I}_{\rm cal}$
\begin{align}\label{formula::PSSAR conformal fitted residual}
\widehat{\epsilon}_i^{\rm cal} = y_i\ominus\widehat{\mu}_c^{\rm train} - \widehat{q}_{\rm train} - \widehat{\rho}_{\rm train} \sum_{j=1}^n w_{i,j} (y_j\ominus\widehat{\mu}_c^{\rm train}- \widehat{q}_{\rm train})  
\end{align}
with $\widehat{q}_{\rm train}=\{{\rm card}(\mathcal{I}_{\rm train})\}^{-1}\sum_{i\in\mathcal{I}_{\rm train}}y_i\ominus\widehat{\mu}_c^{\rm train}$. Let $\widehat{S}_{\alpha,{\rm PSSAR}}$ be the $(1-\alpha)$-quantile of $\{\widehat{r}_i\}_{i\in\mathcal{I}_{\rm cal}}$. The prediction $\widehat{q}_{n+1}^{\rm train}$ using the training set is given by $\widehat{q}_{n+1}^{\rm train} = \widehat{q}_{\rm train} + \widehat{\rho}_{\rm train}  \sum_{j\in \mathcal{I}_{\rm train}}w_{(n+1),j}^*(y_j\ominus\widehat{\mu}_c^{\rm train}-\widehat{q}_{\rm train})$. The prediction set aimed at providing $(1-\alpha)$-level coverage is then constructed as
\begin{align}\label{PSSAR conformal set}
\widehat{\mathcal{Q}}_{\alpha,{\rm PSSAR}} = \left\{y\in\mathcal{S}: \left\|y\ominus\widehat{\mu}_c^{\rm train}-\widehat{q}_{n+1}^{\rm train}\right\|_{\mathcal{C}(\mathcal{H})} \leq \widehat{S}_{\alpha,{\rm PSSAR}}\right\}.
\end{align}

Algorithm~\ref{algorithm::PSSAR conformal} given below summarises the procedure to construct the split-conformal-based prediction set for the PSSAR model.
\begin{algorithm}
\caption{Prediction Set based on PSSAR}\label{algorithm::PSSAR conformal}
\footnotesize
\hspace*{\algorithmicindent} 
\textbf{Input:} Data: $\{y_i\}_{i=1}^{n}$, weight matrix: $W_{n+1}$ containing the spatial information of sites $\{1,2,\ldots,n+1\}$.
  \begin{algorithmic}[1]
    \STATE Randomly split $\{y_i\}_{i=1}^n$ into a training set $\mathcal{Y}_{\rm train}$ and a calibration set $\mathcal{Y}_{\rm cal}$.
    \STATE Fit the PSSAR model in (\ref{model::PSSAR}) using the training data $\mathcal{Y}_{\rm train}$.
    \STATE Calculate the  non-conformity scores $\widehat{r}_i=\|\widehat{\epsilon}_i^{\rm cal}\|_{\mathcal{C} (\mathcal{H})}$ with $\widehat{\epsilon}_i^{\rm cal}$ in (\ref{formula::PSSAR conformal fitted residual}).
    \STATE Form the prediction set for $y_{n+1}$ as $\widehat{\mathcal{Q}}_{\alpha,{\rm PSSAR}}$ defined in (\ref{PSSAR conformal set}).
  \end{algorithmic}
  \hspace*{\algorithmicindent} \textbf{Output:} The $(1-\alpha)$-level prediction set $\widehat{\mathcal{Q}}_{\alpha,{\rm PSSAR}}$ for a future prediction of $y_{n+1}$.
\end{algorithm}

In practice, we consider placing roughly half of the observations in the training set $\mathcal{Y}_{\rm train}$ while the remaining ones are placed in the calibration set $\mathcal{Y}_{\rm cal}$, and this splitting rule is used in all the numerical studies in Sections~\ref{sect::simulation} and \ref{sect::real data}, and Section~S4 in the Supplementary Material. The following theorem shows that the proposed prediction set constructed by Algorithm~\ref{algorithm::PSSAR conformal} is asymptotically~valid.

\begin{theorem}\label{thm::PSSAR conformal}

Under Assumptions~\ref{assump::1}-\ref{assump::4} in the Appendix, for the prediction set $\widehat{\mathcal{Q}}_{\alpha,{\rm PSSAR}}$ defined in~\eqref{PSSAR conformal set}, $\mathbb{P}(y_{n+1}\in \widehat{\mathcal{Q}}_{\alpha,{\rm PSSAR}}) \geq 1-\alpha + o_p(1)$.

\end{theorem}

\section{Spherical Regression Model with Spatial Autoregressive Disturbances}\label{sect::SRMSAR}

In this section, we consider extending the PSSAR model proposed in Section~\ref{sect::PSSAR} by incorporating additional regressors. Returning to the Euclidean setting where $z_1,z_2,\ldots,z_n\in\mathbb{R}$, sometimes the constant zero mean assumption of $z_i$ in different spatial locations is too restrictive. One can extend the Euclidean pure SAR model~\eqref{model::pure SAR process} by introducing real-valued exogenous regressors $\widetilde{x}_i$ as follows (\citealt{anselin1988spatial,lesage2009introduction}):
\begin{align}\label{model::exogenous SAR process}
z_{i}=\widetilde{x}_i^\top\beta+u_i,~u_i=\lambda_0\sum_{i=1}^n w_{i,j}u_j + \varepsilon_i ~i=1,\ldots,n.
\end{align}
This model~\eqref{model::exogenous SAR process} is known as the regression model with spatial autoregressive disturbances, with a vectorised form being
\begin{equation}\label{model::exogenous SAR process vector}
Z_n= X_n\beta + U_n,~U_n=\lambda_0 W_nU_n+\mathcal{E}_n,
\end{equation}
where $Z_n=(z_1,\ldots,z_n)^\top$, $U_n=(u_1,\ldots,u_n)^\top$, $\widetilde{x}_i^\top$ is the $i$\textsuperscript{th} row of $X_n$, and $\mathcal{E}_n=(\varepsilon_1,\ldots,\varepsilon_n)^\top$. Observe that under (\ref{model::exogenous SAR process vector}), one has $\mathbb{E}(Z_n|X_n)=X_n\beta$.

To extend the SAR model with exogenous regressors in (\ref{model::exogenous SAR process vector}) to spherically embedded spatial data, the main challenge is the presence of a different population mean, involving true but unknown parameters $\beta$. For a sequence of spherical observations $y_1,\ldots,y_n\in\mathcal{S}$, if one could obtain the population conditional Fr\'{e}chet mean of each $y_i$, then a similar way to develop the PSSAR model in \eqref{model::PSSAR} can be followed to deal with the regression setting. To tackle this issue, we consider borrowing the idea of the global Fr\'{e}chet regression (\citealt{petersen2019frechet}) to model the conditional Fr\'{e}chet mean of $y_i$ within the SAR framework. For the the regression model with spatial autoregressive disturbances (\ref{model::exogenous SAR process vector}) in the Euclidean setting, the ordinary least-squares estimation of $\beta$ is $\widehat{\beta} = (X_n^\top X_n)^{-1}X_n^\top Z_n.$ This leads to the fitted value of $z_i$, denoted as $\widehat{z}_i$, given by
\begin{align*}
\widehat{z}_i &= \widetilde{x}_i^\top (X_n^\top X_n)^{-1}X_n^\top  Z_n  =  n^{-1}\sum_{j=1}^n \left\{ 1+ (x_i-\widehat{x})\widehat{\Sigma}^{-1}(x_j-\widehat{x}) \right\}z_j \\
     & = \argmin_{\nu\in\mathbb{R}} n^{-1}\sum_{j=1}^n \left\{ 1+ (x_i-\widehat{x})\widehat{\Sigma}^{-1}(x_j-\widehat{x}) \right\} d_E^2(\nu,z_i),
\end{align*}
where $\widetilde{x}_i=(1,x_i^\top)^\top$, $\widehat{x}=n^{-1}\sum_{i=1}^n x_i$, $\widehat{\Sigma}=n^{-1}\sum_{i=1}^n (x_i-\widehat{x})(x_i-\widehat{x})^\top$, and $d_E(\cdot,\cdot)$ is the Euclidean distance. That is, the conditional mean of $z_i$ is a weighted mean with the sum of the weights being~1 (\citealt{petersen2019frechet}). The above findings support us to obtain the population conditional Fr\'{e}chet mean of each random object $y_i$ as $\mu_i = \argmin_{\nu\in\mathcal{S}} M(\nu,x_i)$, where $M(\nu,x_i) = \mathbb{E}_{X, Y} \left[\left\{ 1+(x_i-\overline{x})^\top \Sigma^{-1} (X-\overline{x})  \right\}  d_\mathcal{S}^2(\nu, Y) \right]$, $\overline{x}=\mathbb{E}(x_1)$, and $\Sigma={\rm Cov}(x_1)$ and denote the space of $\{x_i\}_{i=1}^n$ as $\mathcal{X}$ which is a compact subset of the Euclidean space. We can then develop the spherical regression model with spatial autoregressive disturbances (SRMSAR) model as
\begin{align}\label{model::SRMSAR}
\xi_{i}-\overline{\xi}=\lambda_{0}\sum_{i=1}^n w_{i,j}(\xi_j-\overline{\xi}) +\varepsilon_i,~i=1,\ldots,n,
\end{align}
where $\xi_i=y_i \ominus \mu_i$, $\overline{\xi}=\mathbb{E}[\xi_1]$, and $\{\varepsilon_i\}_{i=1}^n\subset \mathcal{C}(\mathcal{H})$ are IID random noise with mean 0 and finite fourth moments. 

\subsection{Estimation and Hypothesis Testing}\label{subsect::SRMSAR est and test}

In practice, the population conditional Fr\'{e}chet means $\mu_1,\mu_2,\ldots,\mu_n$ are unknown and a consistent estimator for each $i=1,2,\ldots,n$ can be obtained via $\widehat{\mu}_i = \argmin_{\nu\in\mathcal{S}}  n^{-1}\sum_{j=1}^n \Big\{ 1+ (x_i-\widehat{x})\widehat{\Sigma}^{-1}(x_j-\widehat{x}) \Big\} d_\mathcal{S}^2(\nu,y_i)$, with the asymptotic behaviour given by the following proposition. 

\begin{proposition}\label{prop::consistency of Frechet mean}
Under Assumptions~\ref{assump::1}-\ref{assump::3} in the Appendix, for observations $y_1,y_2,\ldots,y_n$ follow the model (\ref{model::SRMSAR}), $\sup_{1\leq i\leq n}\sqrt{n\eta_n^{-1}}d_\mathcal{S}(\widehat{\mu}_i,\mu_i)=O_p(1)$.
\end{proposition}

Proposition~\ref{prop::consistency of Frechet mean} allows us to derive the estimation theory of the SRMSAR model. For $i=1,2,\ldots,n$, let $\widehat{\xi}_i=y_i\ominus \widehat{\mu}_i$ and $\widehat{\xi}=n^{-1}\sum_{i=1}^n\widehat{\xi}_i$. Following Proposition~\ref{prop::MoM construct}, the GMM estimator $\widehat{\lambda}$ is 
\begin{equation}\label{formula::SRMSAR lambda est}
\widehat{\lambda} = \argmin_{\lambda} \left[ {\rm tr}\left\{ S_n^\top(\lambda) P_n S_n(\lambda) \widehat{D}_n \right\} \right]^2 ,   
\end{equation}
where the $(j,k)$\textsuperscript{th} element of $\widehat{D}_n$ is $\langle \widehat{\xi}_j-\widehat{\xi} , \widehat{\xi}_k-\widehat{\xi} \rangle_{\mathcal{C}(\mathcal{H})}$. The next theorem below establishes the limiting distribution of the parameter estimation $\widehat{\lambda}$.

\begin{theorem}\label{thm::SRMSAR lambda est}
Under Assumptions~\ref{assump::1}--\ref{assump::3} in the Appendix, assume that $\lim_{n\to\infty}n^{-1}\eta_n {\rm tr} \{(P_n+P_n^\top)W_nS_n^{-1}(\lambda_0) \}\neq 0$. The estimator $\widehat{\lambda}$ from~\eqref{formula::SRMSAR lambda est} has the asymptotic behaviour as $\sqrt{n\eta_n^{-1}} (\widehat{\lambda}-\lambda_0)\xrightarrow{d} N(0,\sigma_\lambda^2)$, where 
\[
\sigma_\lambda^2 = \lim_{n\to\infty}\left[ \frac{\varkappa_4{\rm tr}\left\{ P_n(P_n+P_n^\top)\right\}+(\varkappa_5-2\varkappa_4-1)\sum_{i=1}^n P_{n,ii}^2}{\eta_nn^{-1}{\rm tr}^2\left\{ (P_n+P_n^\top)W_nS_n^{-1}(\lambda_0) \right\}} \right],
\]
$\varkappa_1={\rm tr}(\Sigma_\varepsilon)$, $\Sigma_\varepsilon=\mathbb{E}( \varepsilon_1\circ\varepsilon_1 )$, $\varkappa_2=\mathbb{E}\left( \langle \varepsilon_1,\varepsilon_1\rangle_{\mathcal{C}(\mathcal{H})}^2 \right)$, $\varkappa_3={\rm tr}(\Sigma_\varepsilon^2)$, $\varkappa_4=\varkappa_3/\varkappa_1^2$, and $\varkappa_5=\varkappa_2/\varkappa_1^2$. If we further have 
\begin{inparaenum}
\item[(i)] $P_{n,ii}=0$ for all $i=1,\ldots,n$, or 
\item[(ii)] $\lim_{n\to\infty}\eta_n\to\infty$,
\[
\sigma_\lambda^2 = \lim_{n\to\infty}\left[ \frac{\varkappa_4{\rm tr}\left\{ P_n(P_n+P_n^\top)\right\}}{\eta_nn^{-1}{\rm tr}^2\left\{ (P_n+P_n^\top)W_nS_n^{-1}(\lambda_0) \right\}} \right].
\]
\end{inparaenum}
\end{theorem}

We now consider testing the presence of the spatial effect with the following null and alternative hypotheses
\begin{align}\label{two sample hypothesis SRMSAR}
{\rm H}_0: \lambda=0 ~~{\rm vs.}~~{\rm H}_a:\lambda\neq0.
\end{align}
A Wald-type test statistic for test ${\rm H}_0$ is $\widehat{T}_{w,\lambda} = n\eta_n^{-1}\widehat{\lambda}^2/\widehat{\sigma}_\lambda^2$, where $\widehat{\sigma}_\lambda^2$ is a consistent estimator of $\sigma_\lambda^2$ in Theorem~\ref{thm::SRMSAR lambda est}, and takes the form of
\[
\widehat{\sigma}_\lambda^2 = \frac{\widehat{\varkappa}_4{\rm tr}\left\{ P_n(P_n+P_n^\top)\right\}+(\widehat{\varkappa}_5-2\widehat{\varkappa}_4-1)\sum_{i=1}^n P_{n,ii}^2}{\eta_nn^{-1}{\rm tr}^2\left\{ (P_n+P_n^\top)W_nS_n^{-1}(\widehat{\lambda}) \right\}}, 
\]
with $\widehat{\varkappa}_1={\rm tr}(\widehat{\Sigma}_\varepsilon)$, $\widehat{\Sigma}_\varepsilon=n^{-1}\sum_{i=1}^n \widehat{\varepsilon}_i\circ\widehat{\varepsilon}_i$, $\widehat{\varepsilon}_i=\widehat{\xi}_i-\widehat{\xi}-\widehat{\lambda}\sum_{j=1}^n w_{i,j}(\widehat{\xi}_j-\widehat{\xi})$ $i=1,\ldots,n$, $\widehat{\varkappa}_2=n^{-1}\sum_{i=1}^n \langle \widehat{\varepsilon}_i,\widehat{\varepsilon}_i\rangle_{\mathcal{C}(\mathcal{H})}^2$, $\widehat{\varkappa}_3=\{n(n-1)\}^{-1}\sum_{i\neq j} \langle \widehat{\varepsilon}_i ,\widehat{\varepsilon}_j \rangle_{\mathcal{C}(\mathcal{H})}$, $\widehat{\varkappa}_4=\widehat{\varkappa}_3/\widehat{\varkappa}_1^2$, and $\widehat{\varkappa}_5=\widehat{\varkappa}_2/\widehat{\varkappa}_1^2$. If we further have (i) $P_{n,ii}=0$ for all $i=1,\ldots,n$, or (ii) $\lim_{n\to\infty}\eta_n\to\infty$,
\[
\widehat{\sigma}_\lambda^2 = \left[ \frac{\widehat{\varkappa}_4{\rm tr}\left\{ P_n(P_n+P_n^\top)\right\}}{\eta_nn^{-1}{\rm tr}^2\left\{ (P_n+P_n^\top)W_nS_n^{-1}(\widehat{\lambda}) \right\}} \right].
\] 

\begin{theorem}\label{thm::SRMSAR wald test}
Under Assumptions~\ref{assump::1}--\ref{assump::3} in the Appendix, assume that $\lim_{n\to\infty}n^{-1}\eta_n {\rm tr} \{(P_n+P_n^\top)W_nS_n^{-1}(\lambda_0)\}\neq 0$. Then, under the null hypothesis ${\rm H}_0$ in (\ref{two sample hypothesis SRMSAR}), $\widehat{T}_{w,\lambda}\xrightarrow{d}\chi^2(1)$ as $n\to\infty$.
\end{theorem}

The proof of Theorem~\ref{thm::SRMSAR wald test} is similar to that of Theorem~\ref{thm::wald test} and is omitted. As with the PSSAR model, for a small sample size $n$, the estimate $\widehat{\sigma}_\lambda^2$ may suffer from the curse of dimensionality. An analogous bootstrap-based procedure for the SRMSAR model is:
\begin{itemize}
    \vspace{-0.8em}
    \item[(a)] For $i=1,2,\ldots,n$, calculate $\widehat{\varepsilon}_i=\widehat{\xi}_i-\widehat{\xi}$;
    \vspace{-0.8em}
    \item[(b)] Draw a bootstrap sample $\{\widehat{\varepsilon}_i^*\}_{i=1}^n$ from $\{\widehat{\varepsilon}_i\}_{i=1}^n$ and get $\widehat{\xi}_i^*=\widehat{\varepsilon}_i^*+\widehat{\xi}$;
    \vspace{-0.8em}
    \item[(c)] Calculate the bootstrap estimation $\widehat{\lambda}^{(b)}$ based on $\{\widehat{\xi}_i^*\}_{i=1}^n$;
    \vspace{-0.8em}
    \item[(d)] Repeat (a)-(c) $B$ times to obtain the empirical distribution function of $\{\widehat{\lambda}^{(b)}\}_{b=1}^B$. For a nominal level $\alpha$, if $\widehat{\lambda}$ locates outside the $(1-\alpha)$-level confidence interval based on the empirical distribution function of $\{\widehat{\lambda}^{(b)}\}_{b=1}^B$, denoted as $[l_{\alpha/2}^{\rm PSSAR},u_{1-\alpha/2}^{\rm SRMSAR}]$, reject the null hypothesis ${\rm H}_0$.
\end{itemize}

The following theorem ensures the validity of the proposed bootstrap procedure.
\begin{theorem}\label{thm::SRMSAR bootstrap}
Under Assumptions~\ref{assump::1} and~\ref{assump::3} in the Appendix, assume that $\lim_{n\to\infty}n^{-1}\eta_n {\rm tr} \{(P_n+P_n^\top)W_nS_n^{-1}(\rho_0)\}\neq 0$. Then, under the null hypothesis ${\rm H}_0$ in (\ref{two sample hypothesis SRMSAR}), $\lim_{n \to \infty} \mathbb{P}\left( l_{\alpha/2}^{\rm SRMSAR} \leq \widehat{\rho} \leq u_{1-\alpha/2}^{\rm SRMSAR} \right) = 1-\alpha$.
\end{theorem}

The proof of Theorem~\ref{thm::SRMSAR bootstrap} is similar to that of Theorem~\ref{thm::PSSAR bootstrap} and is omitted.

\subsection{Prediction and Conformal Prediction Set}\label{subsect::SRMSAR prediction}

For the SRMSAR model in (\ref{model::SRMSAR}) with observations $y_1,y_2,\ldots,y_n$ and the weight matrix $W_n$, given the estimated parameters $\{\widehat{\mu}_i\}_{i=1}^n$ $\widehat{\xi}$, and $\widehat{\lambda}$, the prediction for a new skew-symmetric operator at site $n+1$ with the weight matrix $W_{n+1}^*$ is $\widehat{\xi}_{n+1}=\widehat{\xi} + \widehat{\lambda} \sum_{j=1}^{n} w_{n+1,j}^* (\widehat{\xi}_j-\widehat{\xi})$. The prediction for the spherical data $y_{n+1}$ is then $\widehat{y}_{n+1} = \exp(\widehat{\xi}_{n+1})\widehat{\mu}_{n+1} =\left[{\rm id}+\sin(\widehat{\theta})\widehat{\xi}_{n+1}+\{1-\cos(\widehat{\theta})\}\widehat{\xi}_{n+1}^2\right]\widehat{\mu}_{n+1}$, where $\widehat{\theta}=\|\widehat{\xi}_{n+1}\widehat{\mu}_{n+1}\|_\mathcal{H}$ and $\widehat{\mu}_{n+1} = \argmin_{\nu\in\mathcal{S}}  n^{-1}\sum_{j=1}^n \left\{ 1+ (x_{n+1}-\widehat{x})\widehat{\Sigma}^{-1}(x_j-\widehat{x}) \right\} d_\mathcal{S}^2(\nu,y_i)$ with $x_{n+1}$ being the corresponding covariate for $y_{n+1}$.

To quantify the uncertainty of the prediction, we follow a similar idea proposed in Section~\ref{subsect::PSSAR prediction}. We first randomly split $\{(y_i,x_i)\}_{i=1}^n$ into a training set $\mathcal{D}_{\rm train}$ and a calibration set $\mathcal{D}_{\rm cal}$ with the corresponding weight matrices. For the training set, we train the SRMSAR model in (\ref{model::SRMSAR}) using $\{\widehat{\xi}_k^{\rm train}=y_k\ominus\widehat{\mu}_i^{\rm train}:k\in \mathcal{I}_{\rm train}\}$ where $\widehat{\mu}_i^{\rm train}=\argmin_{\nu\in\mathcal{S}}\frac{1}{{\rm card}(\mathcal{I}_{\rm train})}\sum_{k\in \mathcal{I}_{\rm train}}\left\{ 1+ (x_i-\widehat{x}_{\rm train})\widehat{\Sigma}_{\rm train}^{-1}(x_k-\widehat{x}_{\rm train}) \right\} d_\mathcal{S}^2(\nu,y_k)$ with $\widehat{x}_{\rm train}=\{{\rm card}(\mathcal{I}_{\rm train})\}^{-1}\sum_{k\in \mathcal{I}_{\rm train}} x_k$, $\widehat{\Sigma}_{\rm train}=\{{\rm card}(\mathcal{I}_{\rm train})\}^{-1}\sum_{k\in \mathcal{I}_{\rm train}} (x_k-\widehat{x}_{\rm train})(x_k-\widehat{x}_{\rm train})^\top$, and the estimated spatial parameter is denoted as $\widehat{\lambda}_{\rm train}$. The non-conformity score is calculated as $\widehat{R}_i=\|\widehat{\varepsilon}_i^{\rm cal}\|_{\mathcal{C} (\mathcal{H})}$ where for $i\in\mathcal{I}_{\rm cal}$
\begin{align}\label{formula::SRMSAR conformal fitted residual}
\widehat{\varepsilon}_i^{\rm cal} = y_i\ominus\widehat{\mu}_i^{\rm train} - \widehat{\xi}_{\rm train} - \widehat{\lambda}_{\rm train} \sum_{j=1}^n w_{i,j} (y_j\ominus\widehat{\mu}_c^{\rm train}- \widehat{\xi}_{\rm train})  
\end{align}
with $\widehat{\xi}_{\rm train}=\{{\rm card}(\mathcal{I}_{\rm train})\}^{-1}\sum_{i\in\mathcal{I}_{\rm train}}y_i\ominus\widehat{\mu}_i^{\rm train}$. Let $\widehat{S}_{\alpha,{\rm SRMSAR}}$ be the $(1-\alpha)$-quantile of $\{\widehat{R}_i\}_{i\in\mathcal{I}_{\rm cal}}$. The prediction $\widehat{\xi}_{n+1}^{\rm train}$ is given by $\widehat{\xi}_{n+1}^{\rm train} = \widehat{\xi}_{\rm train} + \widehat{\lambda}_{\rm train}  \sum_{j\in \mathcal{I}_{\rm train}}w_{(n+1),j}^*(y_j\ominus\widehat{\mu}_j^{\rm train}-\widehat{\xi}_{\rm train} )$. The prediction set aiming to provide a $(1-\alpha)$-level coverage is then constructed as 
\begin{align}\label{SRMSAR conformal set}
    \widehat{\mathcal{Q}}_{\alpha,{\rm SRMSAR}} = \left\{ y\in\mathcal{S}: \left\|y\ominus\widehat{\mu}_{n+1}^{\rm train}-\widehat{\xi}_{n+1}^{\rm train}\right\|_{\mathcal{C}(\mathcal{H})} \leq \widehat{S}_{\alpha,{\rm SRMSAR}}  \right\},
\end{align}
where $\widehat{\mu}_{n+1}^{\rm train}=\argmin_{\nu\in\mathcal{S}}\{{\rm card}(\mathcal{I}_{\rm train})\}^{-1}\sum_{k\in \mathcal{I}_{\rm train}}\Big\{ 1+ (x_{n+1}-\widehat{x}_{\rm train})\widehat{\Sigma}_{\rm train}^{-1}(x_k-\widehat{x}_{\rm train}) \Big\} d_\mathcal{S}^2(\nu,y_k)$.

Algorithm~\ref{algorithm::SRMSAR conformal} below provides the procedure for constructing the split-conformal-based prediction set for the SRMSAR model, with a theoretical guarantee provided by Theorem~\ref{thm::SRMSAR conformal}.  The same splitting rule as the one introduced in Section~\ref{subsect::PSSAR prediction} is used in the numerical studies in Section~S5.2 in the Supplementary Material.
\begin{algorithm}
\caption{Prediction Set based on SRMSAR} 
  \label{algorithm::SRMSAR conformal}
  \footnotesize
  \hspace*{\algorithmicindent} \textbf{Input:} Data: $\{(y_i,x_i)\}_{i=1}^n$, weight matrix: $W_{n+1}$ containing the spatial information of sites $\{1,2,\ldots,n+1\}$, covariate information for the prediction: $x_{n+1}$.
  \begin{algorithmic}[1]
    \STATE Randomly split $\{(y_i,x_i)\}_{i=1}^n$ into a training set $\mathcal{D}_{\rm train}$ and a calibration set $\mathcal{D}_{\rm cal}$.
    \STATE Fit the SRMSAR model in (\ref{model::SRMSAR}) using the training data $\mathcal{D}_{\rm train}$.
    \STATE Calculate the  non-conformity scores $\widehat{R}_i=\|\widehat{\varepsilon}_i^{\rm cal}\|_{\mathcal{C} (\mathcal{H})}$ with $\widehat{\varepsilon}_i^{\rm cal}$ in (\ref{formula::SRMSAR conformal fitted residual}).
    \STATE Form the prediction set for $y_{n+1}$ as $\widehat{\mathcal{Q}}_{\alpha,{\rm SRMSAR}}$ defined in (\ref{SRMSAR conformal set}).
  \end{algorithmic}
  \hspace*{\algorithmicindent} \textbf{Output:} The $(1-\alpha)$-level prediction set $\widehat{\mathcal{Q}}_{\alpha,{\rm SRMSAR}}$ for a future prediction of $y_{n+1}$.
\end{algorithm}

\begin{theorem}\label{thm::SRMSAR conformal}
Under Assumptions~\ref{assump::1}-\ref{assump::4} in the Appendix, for the prediction set $\widehat{\mathcal{Q}}_{\alpha,{\rm SRMSAR}}$ defined in~\eqref{SRMSAR conformal set}, $\mathbb{P}(y_{n+1}\in \widehat{\mathcal{Q}}_{\alpha,{\rm SRMSAR}}|x_{n+1}) \geq 1-\alpha + o_p(1)$.
\end{theorem}

\section{Simulation Studies}\label{sect::simulation}

We illustrate our method by the simulation of spherically embedded spatial data with a constant Fr\'{e}chet mean for each observation, and we consider the number of neighbours to stay fixed as the sample size grows. Additional simulation studies on the increasing number of neighbours as the sample size grows and on spherically embedded spatial data with additional covariates can be found in Sections~S4 and S5 of the Supplementary Material, respectively. For spherically embedded spatial data with a constant Fr\'{e}chet mean, the proposed PSSAR model is evaluated via letting $P_n=W_n$, where the finite-sample performance of the estimation and the inference of the spatial autoregressive parameter $\rho$ is evaluated in Section~\ref{subsect::simulation PSSAR est and inference}. The prediction performance of the PSSAR model is studied in Section~\ref{subsect::simulation PSSAR prediction}. In this section, the data are generated using the PSSAR model in (\ref{model::PSSAR}) and all simulations are conducted over $200$ replications. The evaluation of the performance of our proposed method under model misspecification is presented in Section~S5 of the Supplementary Material.

\subsection{Simulations on Estimation and Inference}\label{subsect::simulation PSSAR est and inference}

For a given sample size $n$, we directly generate the optimal transport data. For $i\in\{1,2,\ldots,n\}$, generate a random error $\epsilon_i=\tilde{\epsilon}_i\ominus \mu_c$ where $\tilde{\epsilon}_i\in\mathcal{S}^{m-1}$ for a specific value of $m\in\{6,111\}$ and is generated IID by the von Mises--Fisher distribution (\citealt{mardia2009directional}) with the mean direction being $\bm{1}_m/\|\bm{1}_m\|_E$ and the concentration parameter being 1. $\bm{1}_m$ denotes a vector of ones with dimension $m$. Based on $\{\epsilon_i\}_{i=1}^n$, we then calculate $\{q_i\}_{i=1}^n$ using (\ref{model::PSSAR}) for a given $\rho_0\in\{-0.7,-0.3,0,0.1,0.4,0.9\}$. The weight matrix $W_n$ in the PSSAR model is constructed as follows. Consider a given number of neighbours, indicated as $k$, for $i=1,2,\ldots,n$ and the $i$\textsuperscript{th} row of the weight matrix $W_n$, randomly select $k$ distinct neighbours from the set $\mathcal{N}_i=\{1,2,\ldots,n\}/\{i\}$, that is, exclude $\{i\}$ from the entire index set $\{1,2,\ldots,n\}$. The $(i,j)$\textsuperscript{th}element of $W_n$ is generated as $w_{i,j}\sim {\rm U}(0,1)$ for $j\in\mathcal{N}_i$ and $w_{i,j}=0$ when $j\notin \mathcal{N}_{i}$. Finally, we standardise the weight matrix by row so that each row sums to one.

For the purpose of assessing the performance of the parameter estimator, we denote $\widehat{\rho}^{(t)}$ as the estimation of $\rho_0$ in the $t$\textsuperscript{th} replicate. The averaged bias of the GMM estimator is ${\rm Bias}=200^{-1}\sum_{t=1}^{200}(\widehat{\rho}^{(t)}-\rho_0)$, and the root mean squared error is $\textrm{RMSE}=\sqrt{\textrm{SD}^2+\textrm{Bias}^2}$ where  ${\rm SD}=\Big\{200^{-1}\sum_{t_1=1}^{200} (\widehat{\rho}^{(t_1)}-200^{-1}\sum_{t_2=1}^{200}\widehat{\rho}^{(t_2)})^2\Big\}^{1/2}$ is the standard deviation of $\widehat{\rho}^{(t)}$. For the metric spaces $\mathcal{S}^5$ and $\mathcal{S}^{110}$, Table~\ref{table::PSSAR estimation} reports the averaged bias and the RMSE of $\widehat{\rho}$ with the fixed neighbour number being 10, the sample size varying in $\{200,500,1000\}$ and $\rho_0\in\{-0.7,-0.3,0,0.1,0.4,0.9\}$. It is seen that, for every fixed $\rho_0$, as the sample size $n$ increases, both the averaged bias and the RMSE decrease. This finding supports our theoretical results in Theorem~\ref{thm::PSSAR rho est}, which show that the GMM estimator $\widehat{\rho}$ is consistent.
\begin{table}[!htb]
\tabcolsep 0.26in
\caption{Bias and RMSE of the generalised method of moments estimation of $\rho$ with the fixed neighbour number being 10, sample size varying in $\{200,500,1000\}$ and $\rho_0\in\{-0.7,-0.3,0,0.1,0.4,0.9\}$ when the metric space is $\mathcal{S}^5$ or $\mathcal{S}^{110}$ .\label{table::PSSAR estimation}}
\centering
\scalebox{0.9}{
\begin{tabular}{@{}c cc | cc | cc | cc @{}}
  \toprule
&&&\multicolumn{2}{c|}{$n=200$}  & \multicolumn{2}{c|}{$n=500$} & \multicolumn{2}{c}{$n=1000$} \\
&Space& $\rho_0$  & Bias  & RMSE &  Bias  & RMSE &Bias  & RMSE \\ 
   \midrule
&&-0.7 & -0.0368 & 0.7047 & -0.0117 & 0.7017 & -0.0087 & 0.7010 \\ 
&&  -0.3 & -0.0410 & 0.3122 & -0.0132 & 0.3044 & -0.0096 & 0.3027 \\ 
&&  0 & -0.0425 & 0.0877 & -0.0137 & 0.0523 & -0.0099 & 0.0409 \\ 
&$\mathcal{S}^5$&  0.1 & -0.0427 & 0.1331 & -0.0138 & 0.1129 & -0.0100 & 0.1080 \\ 
&&  0.4 & -0.0426 & 0.4093 & -0.0137 & 0.4033 & -0.0099 & 0.4020 \\ 
&&  0.9 & -0.0419 & 0.9032 & -0.0131 & 0.9012 & -0.0091 & 0.9007 \\ 
   \hline
&&-0.7 & -0.0346 & 0.7002 & -0.0142 & 0.7001 & -0.0068 & 0.7000 \\ 
&&  -0.3 & -0.0383 & 0.3005 & -0.0158 & 0.3002 & -0.0076 & 0.3001 \\ 
&&  0 & -0.0396 & 0.0182 & -0.0162 & 0.0112 & -0.0078 & 0.0082 \\ 
&$\mathcal{S}^{110}$&  0.1 & -0.0398 & 0.1016 & -0.0163 & 0.1006 & -0.0078 & 0.1003 \\ 
&&  0.4 & -0.0395 & 0.4004 & -0.0161 & 0.4002 & -0.0077 & 0.4001 \\ 
&&  0.9 & -0.0363 & 0.9002 & -0.0147 & 0.9001 & -0.0070 & 0.9000 \\ 
   \bottomrule
\end{tabular}}
\end{table}

When the number of neighbours increases at a slower rate than the sample size $n$, Table~S1 in the Supplementary Material also shows the decreasing trends of the averaged bias and the RMSE as the sample size increases. Here, the number of neighbours is set to be 10 when $n=200$, the number of neighbours equals 20 when $n=500$, while the number of neighbours is 30 when $n=1000$. Moreover, it is worth noting that the decreasing rates of the averaged bias and the RMSE shown in Table~S1 in the Supplementary Material are slower than those observed in Table~\ref{table::PSSAR estimation}. This is because when the sample size increases, the spatial dependence becomes stronger in the setting for Table~S1 in the Supplementary Material. This also supports our theoretical results in Theorem~\ref{thm::PSSAR rho est}, since the convergence rate of $\widehat{\rho}$ depends on the divergence rate of the elements of the weight matrix.

To evaluate the finite-sample performance of the proposed Wald test statistic $\widehat{T}_{w,\rho}$, we assess its power, defined as the percentage of rejections across 200 replicates at the nominal level of $0.05$. Here we consider $\rho_0$ taking values in $\{0,0.1,-0.3,0.4,-0.7,0.9\}$, which corresponds to signal strengths $\{0,1,2,3,4,5\}$. A zero signal strength means that the observations are independent. Otherwise, the spatial dependence is stronger when the signal strength increases. To deal with the possible curse of dimensionality caused by the estimate $\widehat{\sigma}_\rho^2$, a method based on the PCA of the covariance matrix $\widehat{\Sigma}_\epsilon$ is used, compared with the proposed bootstrap-based procedure in Section~\ref{subsect::PSSAR est and test}. The number of the bootstrap replicates in Section~\ref{subsect::PSSAR est and test}, $B$, is set to be 500. As discussed in Section~\ref{subsect::PSSAR est and test}, a threshold based on the fraction of variance explained by the selected eigenvalues needs to be set in practice, and we set it to $90\%$ for all PCA-based methods in the numerical studies.

\begin{figure}[!htb]
\centering
\begin{subfigure}{\linewidth}
\centering
\includegraphics[width=\linewidth]{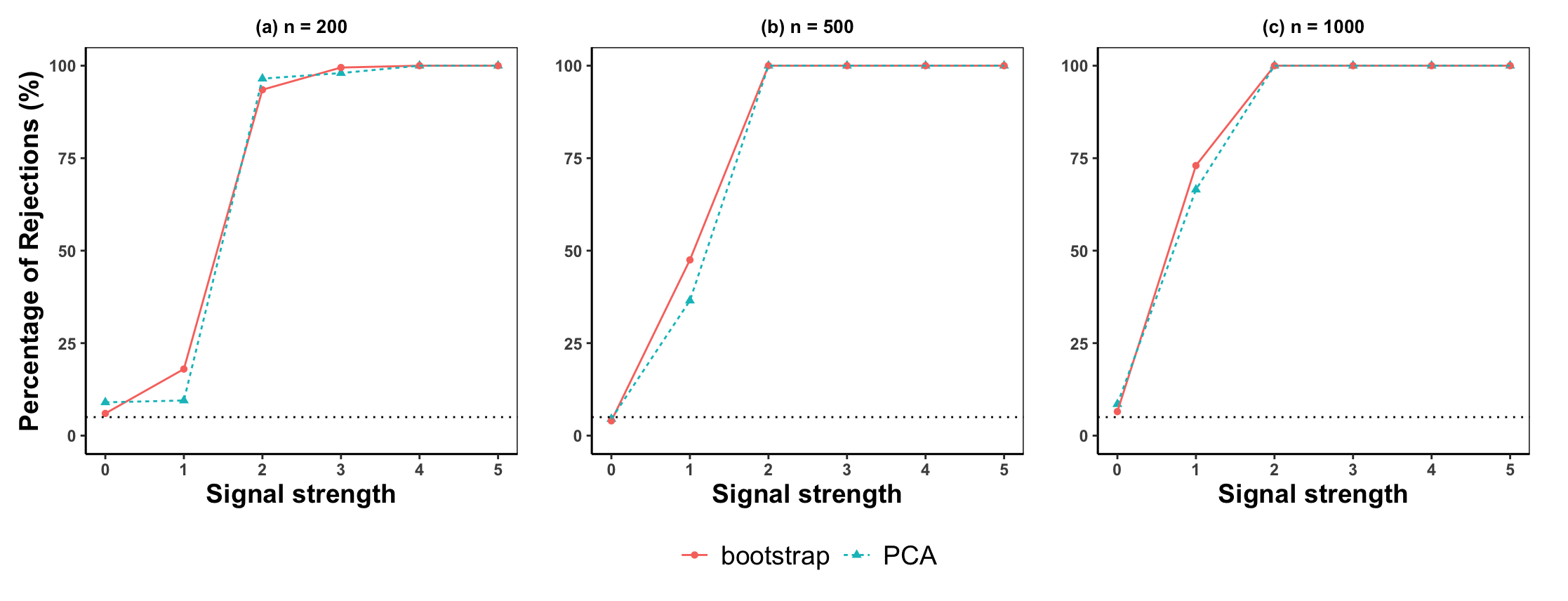}
\end{subfigure}
\vspace{0.5em} 
\begin{subfigure}{\linewidth}
\centering
\includegraphics[width=\linewidth]{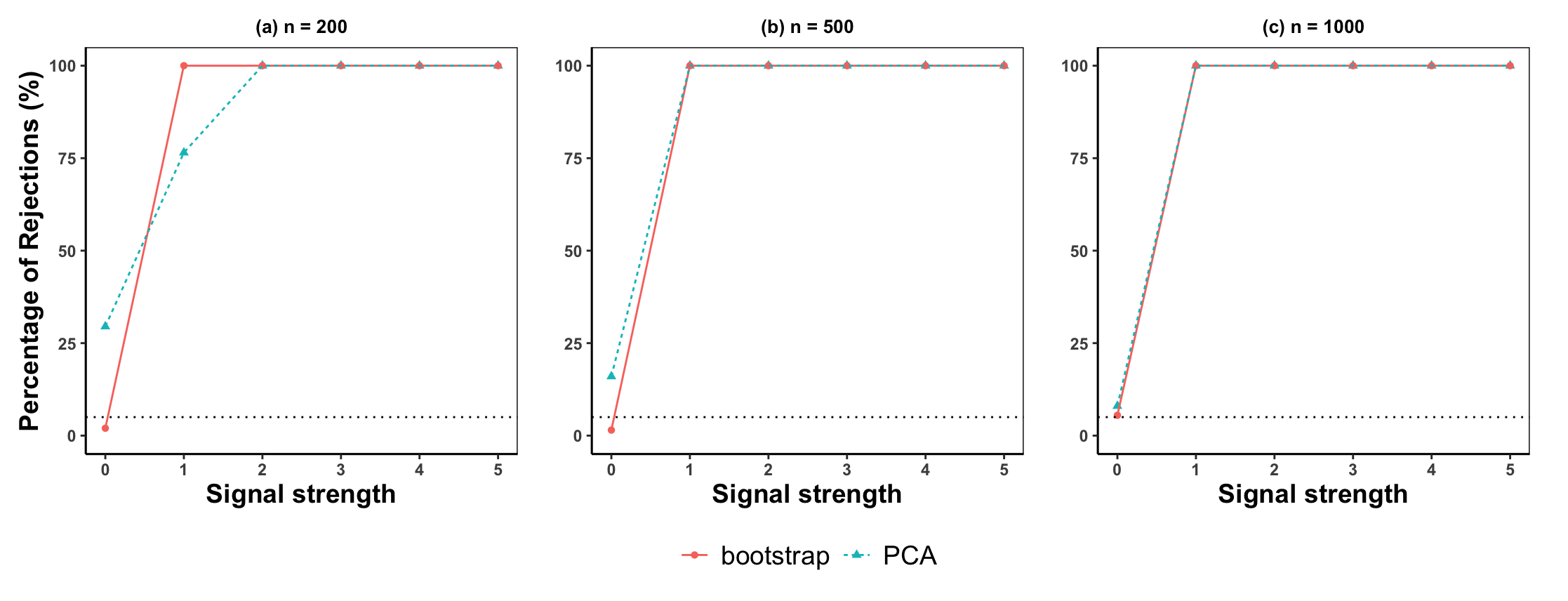}
\end{subfigure}
\caption{The empirical test power of the Wald test based on the PSSAR model using the PCA-based procedure and the bootstrap-based procedure for observations located within $\mathcal{S}^5$ (the first row) or $\mathcal{S}^{110}$ (the second row), with fixed neighbour numbers being 10, signal strengths $\{0,1,2,3,4,5\}$ corresponding to $\rho_0$ taking values in $\{0,0.1,-0.3,0.4,-0.7,0.9\}$, respectively, and sample size varying in (a) $n=200$, (b) $n=500$ and (c) $n=1000$. The blue dashed line represents the PCA-based procedure, and the red solid line represents the bootstrap-based procedure. The horizontal grey dotted line represents the $5\%$ rejection rate.}\label{fig::PSSAR fixed neigh}
\end{figure}

The first row of Figure~\ref{fig::PSSAR fixed neigh} displays the empirical power of the proposed test for observations located within $\mathcal{S}^5$ with the fixed neighbour number being 10 using the PCA-based procedure and the bootstrap-based procedure. As expected, the bootstrap-based procedure outperforms the PCA-based method in most scenarios. The empirical power remains at the significance level $0.05$ under ${\rm H}_0$ when the signal strength is 0, corresponding to $\rho_0=0$, and approaches the significance level as the sample size increases. The empirical power increases as the signal strength grows, and it reaches its maximum value of 100\% faster as the sample size increases. In the second row of Figure~\ref{fig::PSSAR fixed neigh}, the empirical power of the proposed test for observations located within $\mathcal{S}^{110}$ with the fixed neighbour number being 10 using the PCA-based procedure and the bootstrap-based procedure is visualised. As the dimension of $\mathcal{S}^{110}$ becomes much higher, it is not surprising that the PCA-based method performs poorly for a small sample size $n\in\{200,500\}$, especially when the signal strength is 0. When the sample size is larger, $n=1000$, the PCA-based method gives a reasonable result. In contrast, the second row of Figure~\ref{fig::PSSAR fixed neigh} shows the bootstrap-based method maintains good finite-sample performance across all examined sample sizes $n \in \{200, 500, 1000\}$. These findings support the theoretical results of Theorems~\ref{thm::wald test} and \ref{thm::PSSAR bootstrap}.

Figure~S1 in the Supplementary Material reports the results when the number of neighbours increases at a slower rate than the sample size $n$ and also shows similar patterns to those observed in Figure~\ref{fig::PSSAR fixed neigh}.

\subsection{Simulations on Prediction}\label{subsect::simulation PSSAR prediction}

To evaluate the prediction performance of the proposed PSSAR model, we consider generating optimal transport maps $q_1,q_2,\ldots,q_n$ using the same way as in Section~\ref{subsect::simulation PSSAR est and inference} first, then we obtain $y_1,y_2,\ldots,y_n$ calculated by $q_1,q_2,\ldots,q_n$ based on the mean direction $\bm{1}_m/\|\bm{1}_m\|_E$. Given these observations $y_1,y_2,\ldots,y_n$, we use the leave-one-out framework to evaluate prediction performance. That is, for $t$\textsuperscript{th} replicate, we generate observations $y_1^{(t)},y_2^{(t)},\ldots,y_n^{(t)}$ and use $y_1^{(t)},y_2^{(t)},\ldots,y_{n-1}^{(t)}$ to fit the PSSAR model and predict $y_n^{(t)}$, denoted as $\widehat{y}_n^{(t)}$, using the methodologies in Section~\ref{subsect::PSSAR prediction}. The average prediction error is calculated as $(200)^{-1}\sum_{t=1}^{200}d_\mathcal{S}(\widehat{y}_n^{(t)},y_n^{(t)})$. We also construct the prediction set with the aim of providing a nominal coverage probability varying in $\{95\%,90\%,80\%\}$ by Algorithm~\ref{algorithm::PSSAR conformal} and report the average conditional convergence level, which is calculated as the mean of percentages of the prediction set containing $y_n^{(t)}$ through 200 replicates, and the average width of the prediction set, which is the average value of $\widehat{S}_{\alpha,{\rm PSSAR}}$ obtained via 200 replicates.
\begin{table}[!htb]
\tabcolsep 0.19in
\caption{Average prediction errors, average marginal coverage levels and widths of prediction sets obtained with the PSSAR model and the split-conformal-based prediction set with the target coverage levels varying in $\{95\%,90\%,80\%\}$ and the neighbour number is fixed (10) when the sample size grows. The observations are located within $\mathcal{S}^5$ or $\mathcal{S}^{110}$. \label{table::prediction fix neighour}}
\centering
\scalebox{0.85}{
\begin{tabular}{@{}ccl|ccccc@{}}
  \toprule
&& && & $n$  && \\
& Space & Measure  & $n=200$ &  $n=400$  &$n=600$  & $n=800$ &  $n=1000$ \\ 
   \midrule
& & Average Prediction Error &0.8806 & 0.8491 & 0.8191 & 0.8353 & 0.8649 \\ 
& & $95\%$ Prediction Set Coverage & 0.9450 & 0.9500 & 0.9650 & 0.9450 & 0.9400 \\ 
& & $95\%$ Prediction Set Width & 2.1942 & 2.2413 & 2.1924 & 2.1915 & 2.1932 \\ 
&$\mathcal{S}^5$  &  $90\%$ Prediction Set Coverage &  0.8850 & 0.8950 & 0.9150 & 0.9000 & 0.8750 \\ 
& &  $90\%$ Prediction Set Width & 1.9229 & 1.9547 & 1.9117 & 1.9186 & 1.9175 \\ 
& &  $80\%$ Prediction Set Coverage & 0.7750 & 0.8200 & 0.7950 & 0.8000 & 0.7750 \\ 
& &  $80\%$ Prediction Set Width & 1.6059 & 1.6482 & 1.5919 & 1.6045 & 1.5981 \\ 
\hline
& & Average Prediction Error &  0.5052 & 0.5362 & 0.5429 & 0.5512 & 0.5422 \\ 
& & $95\%$ Prediction Set Coverage &   0.9150 & 0.9250 & 0.9050 & 0.9000 & 0.9200 \\ 
& & $95\%$ Prediction Set Width & 0.9674 & 0.9981 & 1.0007 & 1.0055 & 1.0086 \\ 
&$\mathcal{S}^{110}$ &  $90\%$ Prediction Set Coverage & 0.8750 & 0.8850 & 0.8450 & 0.8400 & 0.8650 \\ 
& &  $90\%$ Prediction Set Width & 0.9063 & 0.9373 & 0.9400 & 0.9463 & 0.9490 \\ 
& &  $80\%$ Prediction Set Coverage &0.7800 & 0.7550 & 0.7350 & 0.7250 & 0.7450 \\ 
& &  $80\%$ Prediction Set Width & 0.8350 & 0.8686 & 0.8703 & 0.8766 & 0.8793 \\ 
\bottomrule
\end{tabular}}
\end{table}

Table~\ref{table::prediction fix neighour} reports the average prediction error, the average prediction set coverage, and the average prediction set width for observations located in $\mathcal{S}^5$ or $\mathcal{S}^{110}$. As we set the number of neighbours to be fixed as the sample size increases, the number of data points used to make a prediction will also remain fixed after the PSSAR model is constructed. This is the reason why we do not observe the average prediction error becomes smaller with the growing sample size in Table~\ref{table::prediction fix neighour}. Table~\ref{table::prediction fix neighour} also shows that the proposed split-conformal-based prediction set provides reasonable quantification of the prediction uncertainty, where the average prediction set coverage remains at the corresponding nominal level in most scenarios. The average prediction set width remains stable when the sample size increases. Most importantly, a finite-sample validity can be observed in Table~\ref{table::prediction fix neighour}, and all the findings support Theorem~\ref{thm::PSSAR conformal} in Section~\ref{subsect::PSSAR prediction}. Similar patterns can be observed in Table~S2 in the Supplementary Material when the number of neighbours increases at a slower rate than the sample size $n$.

To illustrate that ignoring the spherical structure of the data may lead to inferior performance. We then compare the prediction performance of the multivariate SAR (MSAR) model proposed by \cite{zhu2020multivariate} using the simulated data $y_1,y_2,\ldots,y_n$. The prediction produced by the MSAR model, $\tilde{y}_*$, is then projected back to the sphere via $\tilde{y}_*/\|\tilde{y}_*\|_\mathcal{H}$. The source code of the MSAR model is available at \url{https://github.com/XueningZhu/MSAR_code}.

As the dimension $m$ of the sphere increases, the MSAR model requires more parameters to estimate, and the computation becomes infeasible for high dimension $m$. Thus, we consider the sphere with dimension $m=4$ and following the same data generation procedure and using the same leave-one-out framework to evaluate the prediction performance in this section by letting $\rho_0=0.8$ and $n\in\{200,400,600,800,1000\}$. For a sample size $n=200$, the run time for one replicate using the MSAR model is around 33.53 seconds, while the run time for one replicate using the proposed PSSAR model is around 0.09 seconds. The comparison of run times for one fit using the MSAR and PSSAR models with a slightly higher-dimensional sphere is reported in the real data analysis in Section~\ref{subsect::geochemical data}.
\begin{figure}[!htb]
\centering
\includegraphics[width=1\linewidth]{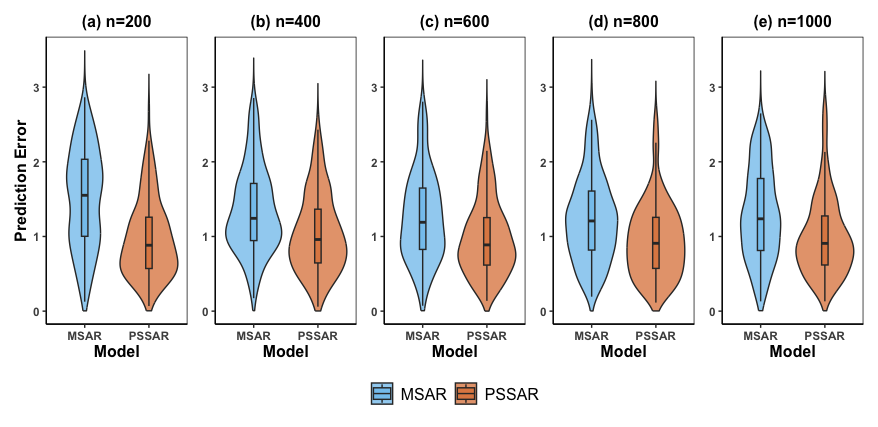}
\caption{Violin plots for the prediction errors produced by the PSSAR model and the MSAR model, with prediction error calculated via angles (radian) between the true observation and the prediction. The simulated data locate in $\mathcal{S}^3$ with sample size $n$ varying in $\{200,400,600,800,1000\}$.}\label{fig:simulation prediction compare}
\end{figure}

Figure~\ref{fig:simulation prediction compare} presents the violin plots for the prediction errors produced by the PSSAR model and the MSAR model, with prediction error calculated via angles (radian) between the true observation and the prediction. From Figure~\ref{fig:simulation prediction compare}, one can see that the proposed PSSAR model consistently outperforms the MSAR model in all scenarios. These simulation results suggest that accounting for the geometric structure of spherically embedded spatial data is beneficial for modelling and prediction. Similar patterns can be observed in Figure~S8 in the Supplementary Material for the SRMSAR model, even under the model misspecification situation.

\section{Empirical Data Analysis}\label{sect::real data}

In this section, we focus on real data analysis on geochemical data in Section~\ref{subsect::geochemical data} and life death counts in Section~\ref{subsect::age distribution data}

\subsection{Geochemical Data in Spain}\label{subsect::geochemical data}

We analyse geochemical composition data generated from agricultural and grazing soils, which consist of measurements of chemical elements in land soils in Europe, sourced from the Geochemical Mapping of Agricultural Soils (GEMAS) project \cite{gosar2014chemistry} and available in the R package \code{robCompositions}. Our study focuses on geochemical compositions in Spain and is restricted to five major chemical elements including Aluminum (Al), Calcium (Ca), Iron (Fe), Potassium (K), and Silicon (Si). The major elements, together with a category containing all the remaining elements, were reported as weight percentage in the collected agricultural soil samples, and this gives compositions located within $\Delta^5$. The number of collected agricultural soil samples is 202, and they were collected from geographically dispersed sites. These site locations satisfied the grid-based setting, and to model spatial dependencies, we adopt a first-order neighbourhood structure. For example, for site $i$ located in the centre of Spain, it has eight neighbours. The $(i,j)$\textsuperscript{th} element of the weight matrix $W$ is then constructed as
\begin{equation*}
w_{i,j} = \frac{\mathcal{I}({\rm site~}j~{\rm is~the~first~order~neighbour~of~site}~i)}{\sum_{j=1}^n \mathcal{I}({\rm site~}j~{\rm is~the~first~order~neighbour~of~site}~i)},
\end{equation*}
where $\mathcal{I}(\cdot)$ is the indicator function. If the site $i$ has no neighbours (e.g., islands), we set $w_{i,j}=0$ for all $j$.

We first use all 202 samples to fit the PSSAR model~\eqref{model::pure SAR process} after the square root transformation for the compositional data. The estimated parameter $\widehat{\rho}=0.676$ with the bootstrap-based test procedure indicates that we can reject the null hypothesis in~\eqref{two sample hypothesis}. 

To evaluate the prediction performance of the proposed PSSAR model, we consider the leave-one-out setting. That is, for every $i=1,2,\ldots,n$ with $n=202$, we use observations excluding $y_i$ to fit the PSSAR model, and, given the fitted model, we predict $y_i$ using (\ref{formula::PSSAR prediction}). Two measures for prediction are considered as 
\begin{inparaenum}
\item[(i)] the great circle distance between the prediction and the true observation calculated as ${\rm Angle}=n^{-1}\sum_{i=1}^n d_\mathcal{S}(\widehat{y}_i,y_i)$, and 
\item[(ii)] the mean squared error (MSE) between the prediction and the true observation ${\rm MSE}=n^{-1}\sum_{i=1}^n \| \widehat{\bm{\delta}}_i - \bm{\delta}_i \|_\mathcal{H}$ using the square transformation as $\bm{\delta}_i=(y_{i1}^2,y_{i2}^2,\ldots,y_{i6}^2)^\top$ where $y_{ij}$ is the element $j$\textsuperscript{th} of $y_i$. 
\end{inparaenum}

\begin{figure}[!htb]
\centering
\includegraphics[width=1\linewidth]{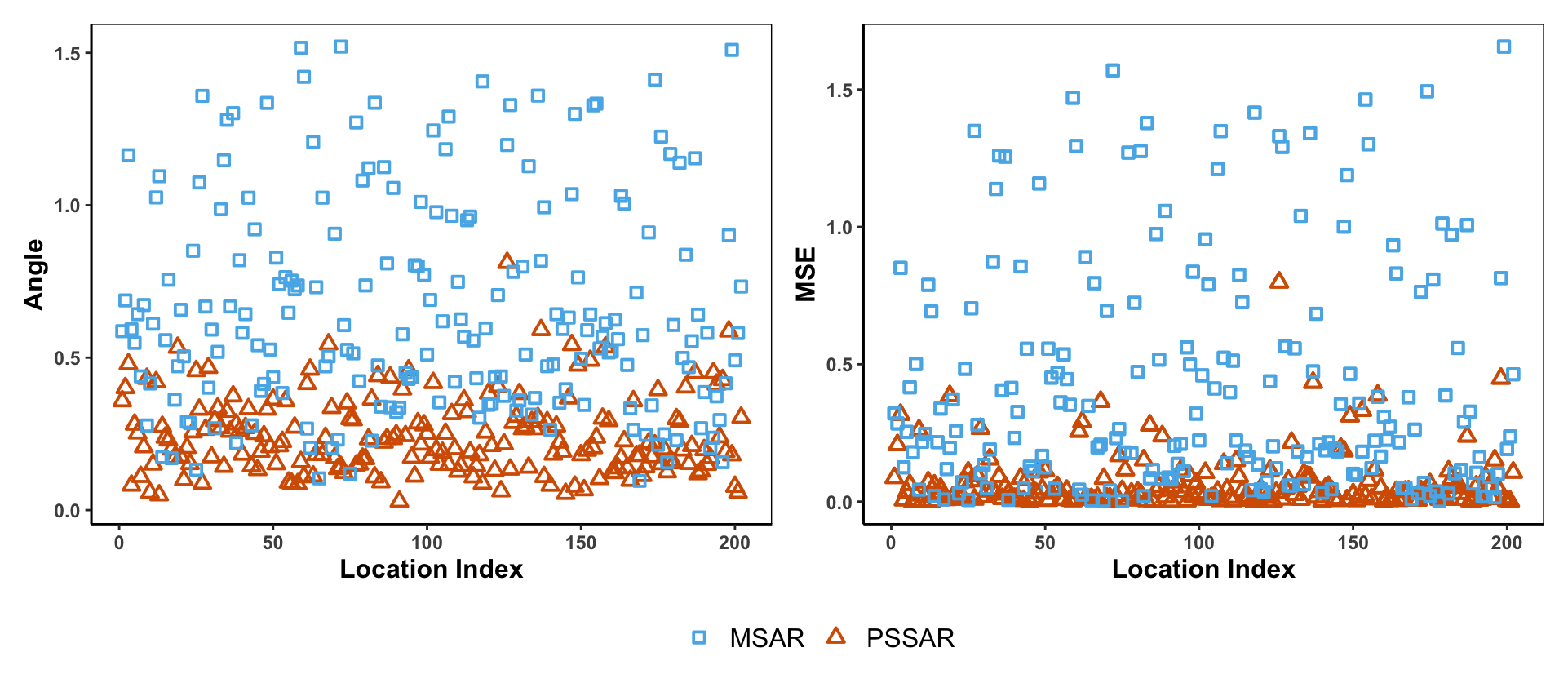}
\caption{Plots for the PSSAR model and the MSAR model with measures produced by (a) angles (radian) between the true observation and the prediction; (b) MSE between the true observation and the prediction using the GEMAS data in Spain.}\label{fig:geochemical prediction dist}
\end{figure}

For comparison, we also consider the MSAR model proposed by \cite{zhu2020multivariate} based on the centre log ratio transformation for the original compositional observations. Notably, as the centre log ratio transformation is undefined for zero values, we adopt the standard practice of replacing zeros with $10^{-0.6}$ and projecting the resulting vectors back onto the simplex \citep{Aitchison1986Compositional}. We then fit the MSAR model to the transformed data, and finally, we apply the inverse centre log ratio transformation to the model predictions to evaluate performance. The run time of one fit using the MSAR model is around 8.15 minutes while the run time of one fit using the PSSAR model is around 0.14 seconds. 

\begin{figure}[!htb]
\centering
\includegraphics[width=1\linewidth]{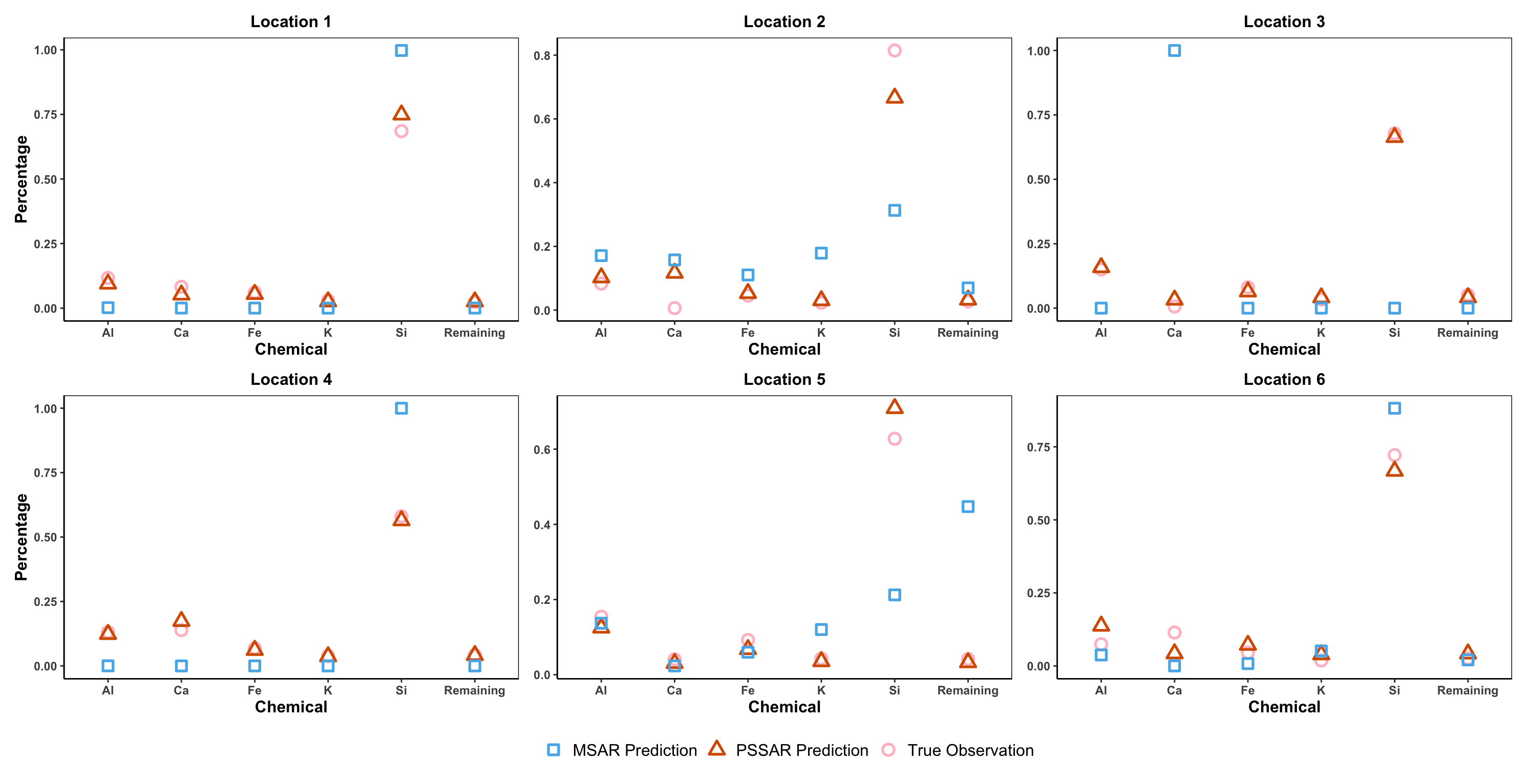}  
\caption{Comparison between the predictions produced by the PSSAR model and the MSAR model and true observations of six randomly selected locations using the GEMAS data in Spain. The predictions produced by the PSSAR model are represented as red triangles, the predictions produced by the MSAR model are represented as blue squares, and the true observations are represented by pink circles.}
\label{fig:geochemical prediction visual}
\end{figure}

Figure~\ref{fig:geochemical prediction dist} compares the two prediction error measures obtained from the proposed PSSAR model and the MSAR model. In most cases, both error measures are substantially smaller for the PSSAR model, indicating its superior predictive performance. To further assess prediction accuracy, we randomly select six sites and present the predicted and true compositional observations in Figure~\ref{fig:geochemical prediction visual}. The PSSAR model produces predictions that closely align with the true observations, whereas the MSAR model produces much worse predictions. Notably, Figure~S9 in the Supplementary Material shows that the variation in the percentage of Ca is significant across the spatial locations. And Figure~\ref{fig:geochemical prediction visual} shows that, across these six locations, the PSSAR model successfully captures the variation in the percentage of Ca in most cases, demonstrating its ability to recover compositional changes given the spatial information.

We now conduct the uncertainty quantification using the proposed split-conformal-based procedure in Section~\ref{subsect::PSSAR prediction}. For the prediction made for site $i$ with $i\in\{1,2,\ldots,202\}$, we evaluate whether it is contained in the 90\% prediction set $\widehat{\mathcal{Q}}_{0.1,{\rm PSSAR}}$ and record the result as $T_i=1$ if the prediction is contained and $T_i=0$ otherwise. The average value of $T_1,T_2,\ldots,T_{202}$ is 85\%, which is closed to the nominal level 90\%.

\begin{figure}[!htb]
\centering
\includegraphics[width=1\linewidth]{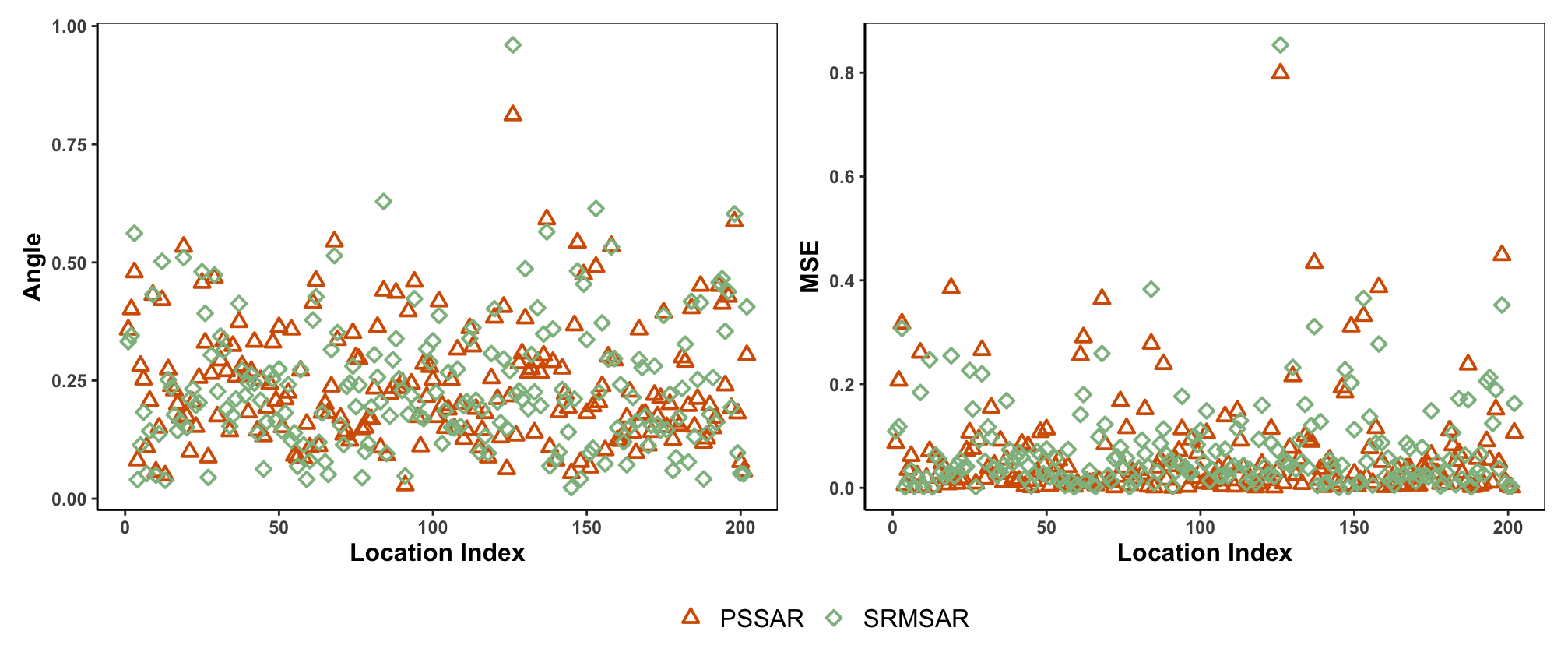}
\caption{Plots for the PSSAR model and the SRMSAR model with measures produced by (a) angles (radian) between the true observation and the prediction; (b) MSE between the true observation and the prediction using the GEMAS data in Spain.}\label{fig:geochemical with covariate prediction dist}
\end{figure}

Finally, we consider using these 202 samples to fit the SRMSAR model in (\ref{model::SRMSAR}) by including three additional covariates: the soil class of the collected agricultural soil samples, the annual mean temperature at the spatial sites, and the annual mean precipitation at the spatial sites. The estimated parameter $\widehat{\lambda}=0.595$ with the bootstrap-based test procedure indicates that we can reject the null hypothesis in~\eqref{two sample hypothesis SRMSAR}. To evaluate the prediction performance of the SRMSAR model, we use the same leave-one-out procedure as for the PSSAR model. Figure~\ref{fig:geochemical with covariate prediction dist} shows the comparison of the two prediction error measures obtained from the proposed PSSAR model and the SRMSAR model. From Figure~\ref{fig:geochemical with covariate prediction dist}, one can observe that the SRMSAR model provides slightly better predictions compared to the one produced by the PSSAR model for most spatial sites. The average great circle distance between the prediction produced via the SRMSAR model and the true observation is around 0.235, which is slightly smaller than the one for the PSSAR model, which is 0.241.

\subsection{Life-Table Death Counts for Japanese Prefectures}\label{subsect::age distribution data}

We use period life-table death counts for Japan, available at \url{https://www.ipss.go.jp/p-toukei/JMD/index-en.asp}. The life-table radix is fixed at 100,000 at age zero, while the number of people alive in the last age group (110+) is zero for each calendar year. There are 111 ages: $0, 1, \dots, 109, 110+$. We work with the probability of dying (i.e., $q_x$) and the radix of the life table to recalculate our estimated death counts (up to six decimal places). In doing so, we obtain more precise life-table death counts ($d_x$) than those reported in \cite{JMD25}. Since the life-table radix is fixed and the number of age groups is 111, the life-table death counts can be treated as the values of a distribution, commonly referred to as the age distribution. We analyse the age distributions of females across 47 Japanese prefectures in the year 2023.

Following \cite{hoshino2024functional}, the $(i,j)$\textsuperscript{th} element of the weight matrix $W_n$ is then constructed as
\begin{equation*}
w_{i,j} = \frac{\mathcal{I}({\rm site~}i~{\rm and}~j~{\rm are~adjacent})\sqrt{{\rm Population_j}}}{\sum_{j=1}^n \mathcal{I}({\rm site~}i~{\rm and}~j~{\rm are~adjacent})\sqrt{{\rm Population_j}}}.
\end{equation*}
When the prefecture $i$ has no neighbours, we set $w_{i,j}=0$ for all $j$. This construction of the weight matrix is based on the assumption that the impacts of demographic changes in large cities should be larger than those in small cities.

Here, we focus on using techniques similar to Granger causality in \cite{shang2021granger} to evaluate whether income affects life-table death counts. We first use the whole 47 samples to fit the PSSAR model (\ref{model::pure SAR process}) after the square root transformation for the distribution data described in Section~\ref{sect::spatial spherical data}. The estimated parameter $\widehat{\rho}=0.408$ with the bootstrap-based test procedure indicates that we can reject the null hypothesis in~\eqref{two sample hypothesis}. We then obtain average income data for every prefecture in 2023 from \url{https://www.e-stat.go.jp/en}, and an SRMSAR model in (\ref{model::SRMSAR}) is fitted by introducing the income data as a covariate. The fitted SRMSAR model gives $\widehat{\lambda}=0.394$ and the bootstrap-based test procedure indicates that we can reject the null hypothesis in~\eqref{two sample hypothesis SRMSAR}. Finally, we evaluate whether including income improves prediction performance.

Utilising the same leave-one-out setting as the one in Section~\ref{subsect::geochemical data}, two measurements of the prediction error are considered. The first one is for the spherical data $\widehat{y}_i,y_i\in\mathcal{S}^\infty$ is the great circle distance between the prediction and the true observation calculated as ${\rm Angle}=n^{-1}\sum_{i=1}^n d_\mathcal{S}(\widehat{y}_i,y_i)$. The second measurement of prediction error is for the distribution data, where we first perform a functional pointwise square transformation as $t_s(y_i)=g_i$ where $g_i(\omega)=y_i^2(\omega)$ for all $\omega\in\mathbb{R}$, and then the Jensen–Shannon divergence between the prediction $\widehat{g}_i=t_s(\widehat{y}_i)$ and the true observation $g_i$ is calculated as
\begin{equation*}
{\rm JSD}=n^{-1}\sum_{i=1}^n \frac{1}{2}{\rm KL}(\widehat{g}_i\|\overline{g}_i) + \frac{1}{2}{\rm KL}(g_i\|\overline{g}_i),
\end{equation*}
where ${\rm KL}(\cdot\|\cdot)$ is the Kullback–Leibler divergence and $\overline{g}_i=(\widehat{g}_i+g_i)/2$ is the mixture distribution of $\widehat{g}_i$ and $g_i$. 
\begin{figure}[!htb]
\centering
\includegraphics[width=1\linewidth]{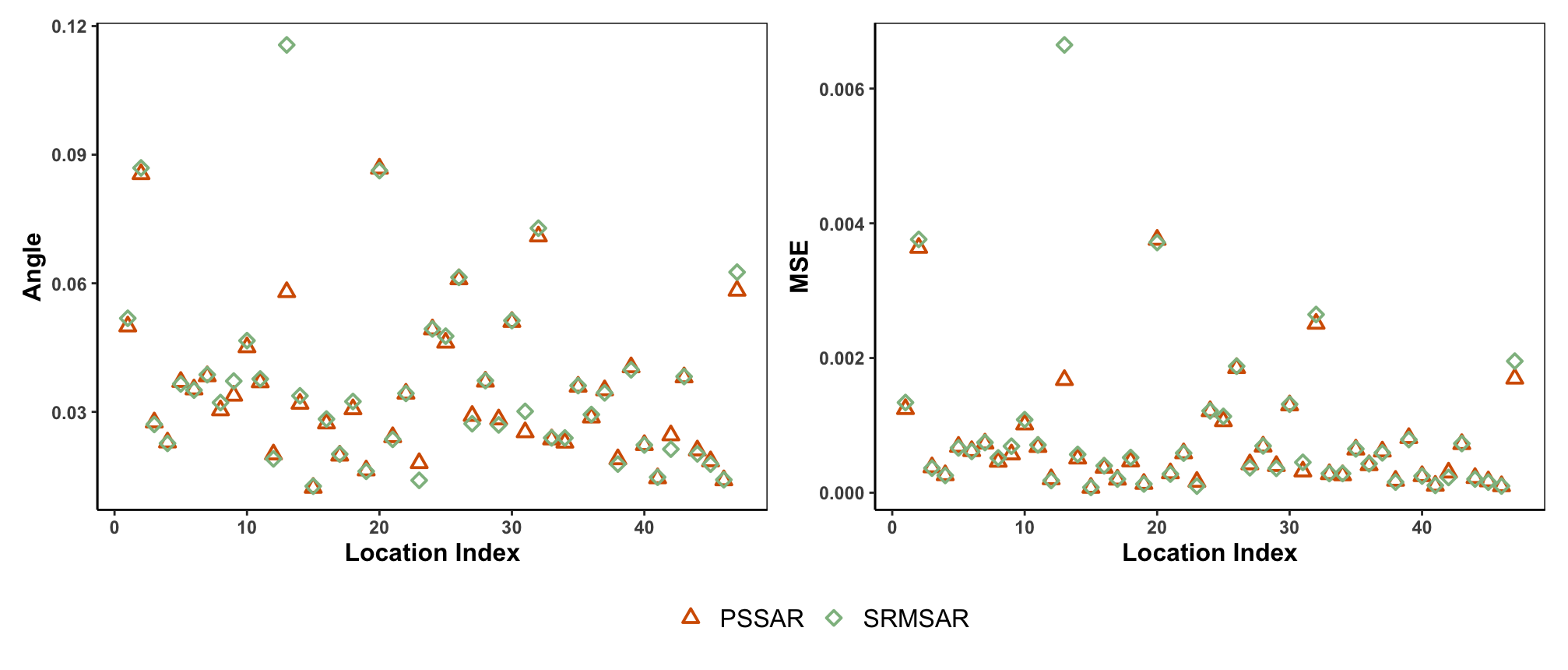}
\caption{Plots for the PSSAR model and the SRMSAR model with measures produced by (a) angles (radian) between the true observation and the prediction; (b) Jensen–Shannon Divergence between the true observation and the prediction using the life-table death count data in Japan.}
\label{fig:LTCD with covariate prediction dist}
\end{figure}

Figure~\ref{fig:LTCD with covariate prediction dist} compares the two prediction error measures obtained from the PSSAR model and the SRMSAR model. The average great circle distance between the prediction produced via the SRMSAR model and the true observation is around 0.036, which is larger than the one for the PSSAR model, which is 0.035. Figure~\ref{fig:LTCD with covariate prediction dist} shows that for most prefectures, including the income data as an additional covariate does not improve the prediction performance. This implies that the income does not Granger-cause the life-table death counts for females in Japan.

\begin{figure}[!htb]
\centering
\includegraphics[width=1\linewidth]{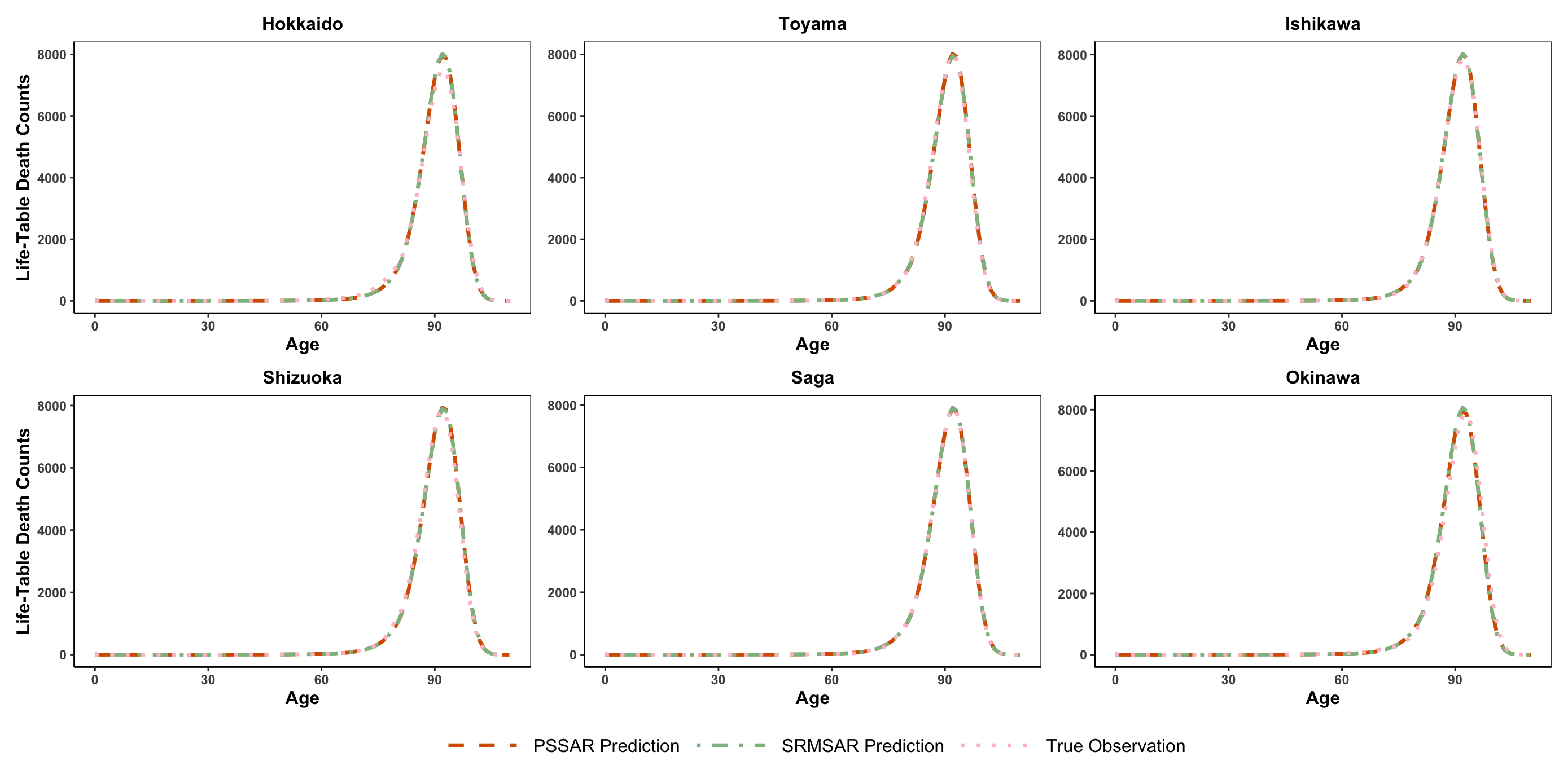}
\caption{Comparison between the predictions produced by the PSSAR model and the SRMSAR model and true observations of six randomly selected prefectures using the life-table death count data in Japan. The predictions from the PSSAR model are shown as red dashed lines, those from the SRMSAR model as green dot-dashed lines, and the true observations as pink dotted lines.}
\label{fig:LTDC prediction visual}
\end{figure}

To further assess prediction accuracy, we randomly select six prefectures and present the predicted and true distributional observations in Figure~\ref{fig:LTDC prediction visual}. One can observe that the predictions from both the PSSAR and SRMSAR models overlap with the true observations, indicating strong predictive accuracy. We then conduct uncertainty quantification using the proposed split-conformal-based procedure in Section~\ref{subsect::PSSAR prediction}, assuming that including income does not improve prediction performance. Following the same procedure as that used in Section~\ref{subsect::geochemical data} to conduct the uncertainty quantification, we get the average value of $T_1, T_2,\ldots, T_{47}$ is 89\%, which is close to the nominal level 90\%.

\section{Conclusion}\label{sec:conclu}

We develop a unified framework for modelling, inference, and prediction for spherically embedded spatial data, a class of object oriented data that is increasingly prevalent in modern scientific applications yet remains underexplored in spatial statistics. By working directly on the sphere and avoiding projections into Euclidean spaces, our approach respects the intrinsic non-Euclidean geometry of the data. It overcomes key limitations of existing methods that rely on restrictive distributional assumptions or transformations.

At the core of the proposed framework is the SSAR model, which extends recent advances in spherical time-series modelling to spatial settings through an optimal transport formulation. The SSAR model accommodates both finite and infinite-dimensional spheres and allows for spatially varying Fr\'{e}chet means via exogenous covariates. We establish a generalised method of moments estimation approach tailored to the spatial dependence structure and develop asymptotic theory. Beyond modelling, we introduce a distribution-free Wald test for assessing spatial dependence in spherically embedded data. The proposed test applies to general Hilbert spheres and does not rely on parametric assumptions, addressing an important gap in the current literature. To mitigate the challenges posed by high dimensionality in finite samples, we further propose a bootstrap-based testing procedure that improves empirical performance while retaining theoretical validity. In addition, we develop a conformal prediction methodology that provides valid predictive uncertainty quantification under spatial dependence, extending conformal inference beyond the IID setting for non-Euclidean data.

The effectiveness and versatility of the proposed framework are demonstrated through extensive simulation studies and real data applications involving geochemical compositions and age distributions of death. These examples illustrate that the SSAR model and the associated inferential tools can capture complex spatial dependence structures while producing accurate predictions in practice.

Several directions for future research remain open. While we focus on extending the regression model with spatial autoregressive disturbances for Euclidean data to deal with the nonconstant Fr\'{e}chet mean setting, another possible way is to allow the spatial dependence parameter affect the Fr\'{e}chet mean. Moreover, while the conditional Fr\'{e}chet mean for spherically embedded spatial data is built upon the idea of the Fr\'{e}chet regression, several extensions of the Fr\'{e}chet regression can be explored into the proposed framework, such as the Fr\'{e}chet concurrent regression in \cite{bhattacharjee2022concurrent}, and the Fr\'{e}chet single index regression in \cite{bhattacharjee2023single} and \cite{ghosal2023frechet}.

\section*{Supplementary Material}

The Supplementary Material includes the deviation of the optimal transport map, several technical lemmas, proofs of the main results, additional simulation settings and results for the PSSAR and SRMSAR models, and an additional figure for empirical data analysis. 

\section*{Acknowledgement}

This research was supported by an Australian Research Council Future Fellowship (FT240100338) and undertaken with assistance from computational resources provided by the Australian Government through the National Computational Infrastructure (NCI) under the Macquarie University Merit Allocation Scheme.

\section*{Appendix}

In this Appendix, we first state the assumptions used to derive our theoretical results. Then, we define the Fr\'{e}chet derivative, which is required in the assumptions.

\begin{assumption}\label{assump::1}
The matrices $W_n$, $S_n^{-1}(\rho_0)$, and $P_n$ are uniformly bounded in both row and column sums. The $(k,l)$\textsuperscript{th} elements of $W_n$ and $P_n$ are of order $O(\eta_n^{-1})$ uniformly in $k$ and $l$, where $\{\eta_n\}$ satisfies $\lim_{n\to\infty}n^{-1}\eta_n=0$.
\end{assumption}

\begin{assumption}\label{assump::2}
The support ${\rm supp}(Y)\subset \mathcal{S}$ of $Y$ satisfy $\sup_{\nu_1,\nu_2\in{\rm supp}(Y)}d(\nu_1,\nu_2)\leq \pi/2$.
\end{assumption}

\begin{assumption}\label{assump::3}
For every $i=1,2,\ldots,n$, the population conditional Fr\'{e}chet mean $\mu_i$ exists and is unique. That is, $\Lambda_i=\lim_{n\to\infty} \mathbb{E}\Big\{n^{-1}\eta_n \sum_{i=1}^n D^2 \overline{M}_n^{(i)}(\nu)\Big\}$ is (continuously) invertible where $\overline{M}_n^{(i)}(\nu)=n^{-1}\sum_{j=1}^n \Big\{ 1+ (x_i-\overline{x})\Sigma^{-1}(x_j-\overline{x}) \Big\} d_\mathcal{S}^2(\nu,y_j)$ for every $i=1,2,\ldots,n$ and $D^2$ is the second-order Fr\'{e}chet derivative of a function. Further assume $\inf_{x\in\mathcal{X}}\inf_{\nu\in\mathcal{S}:d(\nu,\mu(x))>\zeta} \left( M(\nu,x) - M(\mu(x),x) \right)>0$, for any $\zeta>0$ where $\mu(x)=\argmin_{\nu\in\mathcal{S}}M(\nu,x)$. 
\end{assumption}

\begin{assumption}\label{assump::4}
Let $F_{\rm PSSAR}(\tau)=\mathbb{P}\left\{\mathbb{1}(\|\epsilon_1\|_{\mathcal{C}(\mathcal{H})}\leq \tau)\right\}$ and $F_{\rm SRMSAR}(\tau)=\mathbb{P}\left\{\mathbb{1}(\|\varepsilon_1\|_{\mathcal{C}(\mathcal{H})}\leq \tau)\right\}$, assume that $F_{\rm PSSAR}(\tau)$ and $F_{\rm SRMSAR}(\tau)$ are Lipschitz continuous with some constant $L>0$.
\end{assumption}

Assumption~\ref{assump::1} is a regularity condition on the weight matrix $W_n$ in the literature of spatial autoregressive models, while the condition on $P_n$ is standard for the GMM estimation theory under the spatial autoregressive model setting. Notably, in Assumption~\ref{assump::1}, we allow $\eta_n$ diverge to infinity at a rate slower than the rate of the sample size $n$. This works for spatial models with spatial interactions when a spatial observation has a large number of neighbours. Assumption~\ref{assump::2} is needed to ensure the existence and the uniqueness of the Fr\'{e}chet mean of the spherical embedded data, see \cite{dai2022statistical} for more details. Assumption~\ref{assump::3} contains the sufficient condition to ensure the existence and the uniqueness of the conditional Fr\'{e}chet mean of the spherical embedded data and the second part of Assumption~\ref{assump::3} is a uniform population identification condition which is needed to derive the uniform convergence result of the the conditional empirical Fr\'{e}chet mean, see \cite{petersen2019frechet}. Assumption~\ref{assump::4} is used to ensure the validity of the split-conformal-based prediction procedure, which is standard in the conformal prediction literature \citep{zhou2025conformal}.

Below is a definition of the Fr\'{e}chet derivative.
\begin{definition}[\cite{lang2012fundamentals}]
Let $\mathcal{L}$ and $\mathcal{K}$ be Banach spaces with norms $\|\cdot\|_\mathcal{L}$ and $\|\cdot\|_\mathcal{K}$, respectively. A function $f:\overline{\mathcal{L}}\subset \mathcal{L}\mapsto \mathcal{K}$ is Fr\'{e}chet differentiable at a point $x\in \overline{\mathcal{L}}$ if there exists a continuous linear map $g$ of $\mathcal{L}$ onto $\mathcal{K}$ such that for $r\in\mathcal{L}$,
\[
f(x+r)=f(x)+g(r)+\delta(r),
\]
where $\|\delta(r)\|_\mathcal{K\to 0}$ as $\|r\|_\mathcal{L}\to 0$. The linear map $l$ is called the Fr\'{e}chet derivative of $f$ at $x$, denoted as $Df(x)$. If $f$ is Fr\'{e}chet differentiable at every point in $\overline{\mathcal{L}}$, then the derivative $Df$ is a map $Df:\overline{\mathcal{L}}\mapsto \mathcal{B}(\mathcal{L},\mathcal{K})$ where $\mathcal{B}(\mathcal{L},\mathcal{K})$ denotes the space of continuous linear maps from $\mathcal{L}$ into $\mathcal{K}$.
\end{definition}

\bibliographystyle{agsm}
\bibliography{reference}

\end{document}